\documentclass[useAMS,usenatbib]{mn2e} 
\usepackage{amssymb}
\usepackage{times}
\usepackage{multirow}
\usepackage{epsfig}

\newif\ifAMStwofonts

\def\fig#1{Figure~\ref{#1}}
\def\Fig#1{Figure~\ref{#1}}
\def\tab#1{Table~\ref{#1}}

\def\sec#1{Section~\ref{#1}}

\def\appen#1{Appendix~\ref{#1}}

\newcommand{\taulong}[1]{Tau$_{\rm 850}{#1}$}
\newcommand{\taushort}[1]{Tau$_{\rm 450}{#1}$}
\newcommand{\taucso}[1]{Tau$_{\rm CSO}{#1}$}

\newcommand{\apj}{ApJ}

\newcommand{\mnras}{MNRAS}

\newcommand{\aap}{A\&A}

\newcommand{\procspie}{Proc. SPIE}

\title[Atmospheric limitations of submm astronomy]{On the atmospheric
  limitations of ground-based submillimetre astronomy using array receivers} 

\author[E. N. Archibald et al.]{E. N. Archibald$^{1}$\thanks{email:
    e.archibald@jach.hawaii.edu}, T. Jenness$^{1}$, W. S. Holland$^{1,2}$, 
  I. M. Coulson$^{1}$,
  N. E. Jessop$^{1}$,\and J.  A. Stevens$^{2,3}$, E. I. Robson$^{1,4}$,
  R. P. J. Tilanus$^{1,5}$, W. D. Duncan$^{2}$,
    and J. F. Lightfoot$^{2}$\\
  $^1$Joint Astronomy Centre, 660 N. A`oh\={o}k\={u} Place, University
  Park, Hilo, Hawaii, USA \\
  $^2$UK Astronomical Technology Centre, Royal Observatory, Blackford Hill,
  Edinburgh EH9 3HJ, UK \\
  $^3$Mullard Space Science Laboratory, University College London, Holmbury
  St. Mary, Dorking, Surrey, RH5 6NT, UK\\
  $^4$Centre for Astrophysics, University of Central Lancashire, Preston, PR1
  2HE, UK\\
  $^5$Netherlands Organization for Scientific Research (NWO), Postbus 93460, 
       2509 AL Den Haag, the Netherlands\\
  }

\date{Accepted ;
      Received ;
      in original form }

\pagerange{\pageref{firstpage}--\pageref{lastpage}}
\pubyear{2001}

\begin{document} 
\label{firstpage}


\maketitle


\begin{abstract}
  The calibration of ground-based submillimetre observations has always been a
  difficult process.  We discuss how to overcome the limitations imposed by
  the submillimetre atmosphere.  Novel ways to improve line-of-sight opacity
  estimates are presented, resulting in tight relations between opacities at
  different wavelengths.  The submillimetre camera SCUBA, mounted on the JCMT,
  is the first large-scale submillimetre array, and as such is ideal for
  combating the effects of the atmosphere.  For example, we find that the
  off-source pixels are crucial for removing sky-noise.  Benefitting from
  several years of SCUBA operation, a database of deep SCUBA observations has
  been constructed to better understand the nature of sky-noise and the
  effects of the atmosphere on instrument sensitivity.  This has revealed
  several results.  Firstly, there is evidence for positive correlations
  between sky-noise and seeing and sky-noise and sky opacity.  Furthermore,
  850-\micron{} and 450-\micron{} sky-noise are clearly correlated, suggesting
  that 450-\micron{} data may be used to correct 850-\micron{} observations
  for sky-noise.  Perhaps most important of all: if off-source bolometers are
  used for sky-noise removal, there is no correlation between instrument
  sensitivity and chop throw, for chop throws out to 180 arcsec.
  Understanding the effects of submillimetre seeing is also important, and we
  find that the JCMT beam is not significantly broadened by seeing, nor is
  there an obvious correlation between seeing and pointing excursions.
\end{abstract}


\begin{keywords}
submillimetre -- instrumentation: detectors (SCUBA) -- telescopes (James Clerk Maxwell Telescope)
\end{keywords}


\section{Introduction}
\label{intro}

Ground-based observations at submillimetre wavelengths are severely hindered
by the atmosphere, which absorbs, emits, and refracts the incoming radiation.
High, dry sites are needed, such as Mauna Kea in Hawaii.  Even then the
transmission is generally poor, with only a small number of semi-transparent
windows accessible.

To calibrate submillimetre data, the atmospheric opacity must be accurately
determined.  The main absorber of radiation in this waveband is water vapour,
although oxygen and ozone can be significant contributors.  Furthermore, the
transparency of the atmosphere often changes on short timescales.  Thus,
frequent measurements of the opacity are crucial, especially for the shorter
wavelength windows at 350 and 450\,\micron{}.  There is a strong dependence of
the transmission on wavelength; the shorter windows are more opaque, and
deteriorate faster as conditions worsen.

\begin{figure*}
\epsfig{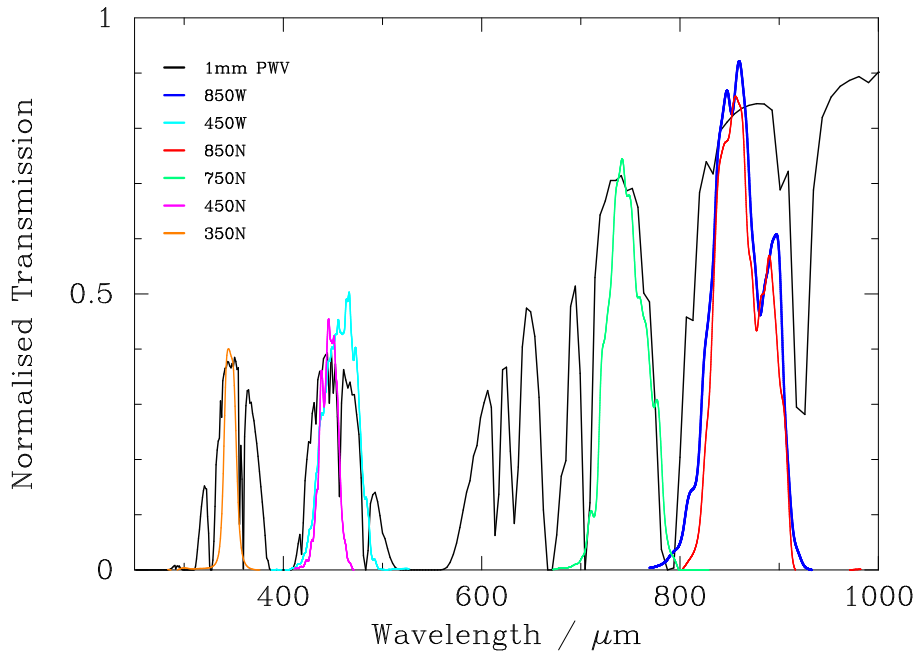}
\caption{The SCUBA filter profiles measured in situ, superimposed on
the submillimetre atmospheric transmission curve for Mauna Kea
assuming 1\,mm of precipitable water vapour.} 
\label{filterfig}
\end{figure*} 

In addition to attenuating the signal, the atmosphere and immediate
surroundings of the telescope emit thermal radiation several orders of
magnitude larger than the source signal. Spatial and temporal variations
in this sky emissivity give rise to "sky noise", which can degrade the
effective instrument sensitivity by up to an order of magnitude.  The
thermal DC offset and sky noise variability can be largely removed by the
conventional techniques of sky-chopping and telescope nodding. To be
effective against sky noise requires the secondary mirror to switch
between sky+source and sky faster than the rate at which the sky is
varying (typically greater than a few Hz). Nodding the telescope primary
to place the source alternately in both chop beams cancels slower varying
sky gradients due to chop-beam imbalances and time-dependent telescope
spillover signals. It is not practical to nod the primary at the chop rate
and so typical frequencies of 0.1 Hz are adopted. However, although these
techniques diminish the effects of sky noise, they do not remove the
residual signature completely.

Fluctuations in the atmospheric refractive index, as the atmosphere drifts
through the telescope beam, cause variations in the path length from source to
telescope \citep{churchhills90,church93}. This is sometimes referred to as
submillimetre seeing, and causes apparent (short-term) pointing shifts, which
ultimately degrade the S/N of an observation. This is particularly problematic
for under-sampled arrays such as SCUBA.  Until the time that adaptive optics
becomes available for the submillimetre, the effects of this seeing must
be closely monitored, and the data severely affected removed from a particular
observation.

This paper considers the limitations imposed on the quality of submillimetre
data by the atmosphere, and presents methods that maximise the accuracy of the
data.  The layout is as follows: \sec{scuba} provides a technical description
of SCUBA and its advantages compared with previous instruments, while
\sec{tau} discusses techniques of determining atmospheric extinction as a
function of wavelength.  We present our own skydip model and consider how
continuously-operating radiometers are crucial for precise measurements of
atmospheric opacity.  Sky-noise is described in detail in \sec{skynoise},
particularly how it can degrade instrument sensitivity, and how most of its
effects can be removed with state-of-the-art instrumentation such as SCUBA.
The effects of refraction noise (submillimetre seeing) are described in
\sec{seeing}.  Using a carefully constructed database of SCUBA observations
and atmospheric information, \sec{sensitivity} analyses how the atmosphere
directly effects instrument sensitivity.  In \sec{future} we conclude with a
discussion of future technology that will provide further improvements in the
accuracy of submillimetre data.

\section{The benefits of array receivers: SCUBA}
\label{scuba}

Until recently, observing in the submillimetre was limited to single-pixel
devices.  UKT14 \citep{ukt14paper}, the forerunner to SCUBA on the JCMT, was
detector-noise limited at all wavelengths of operation, except under periods
of high sky variability, when the sensitivity at 350 and 450 \micron{} was
severely limited by sky-noise.

An array receiver is a major improvement over a single-pixel system.  The
off-source pixels can be used to measure sky-noise on short timescales.  This
is crucial for accurate work: \citet{omc96} compared the performance of a
detector array with a single-pixel system, and found that the poor sky
cancellation offered by a single detector could in fact produce fake
detections of faint sources.

Within the past few years submillimetre astronomy has witnessed the arrival of
the first large format bolometer arrays, together with detector sensitivities
which are limited, under stable conditions, by the background photon noise
from the sky.  The largest and most powerful of this new generation of submm
cameras is the Submillimetre Common-User Bolometer Array \citep{scubapaper},
which operates on the 15-m James Clerk Maxwell Telescope, on Mauna Kea.  SCUBA
is a dual camera system containing 91 pixels in the short-wavelength (SW)
array and 37 pixels in the long-wavelength (LW) array.  Background-limited
performance is achieved by cooling the detectors to $\sim 100$\,mK.  Both
arrays have approximately the same field-of-view (FOV) on the sky (2.3
arcminutes in diameter) and can be used simultaneously by means of a dichroic
beamsplitter.  The SW array is optimised for operation at 450\,\micron{},
while the LW array is optimised for 850\,\micron{}.

The wavelength of operation is selected by a bandpass filter carefully
designed to match the transmission window. The filters are
multi-layer, metal-mesh interference filters \citep{hazellthesis}
located in a nine-position rotating drum that surrounds the
arrays. They have excellent transmission (typically over 80\%), and
also less than 0.1\% out-of-band power leakage. This latter
characteristic is particularly important as it ensures that there is
minimum contribution to the source signal from extraneous sky
emission.  The spectral performance of the filters was measured by the
University of Lethbridge Fourier Transform Spectrometer
\citep{naylorfts}, and the resultant profiles are overlaid on the
Mauna Kea atmospheric transmission curve in \fig{filterfig}.  For a
more detailed look at the submillimetre atmosphere please refer to
\citet{naylorsubatmos}.

SCUBA has three basic observing modes \citep{scubapaper}.  Photometry, uses a
single bolometer to observe point sources.  For sources larger than the beam,
but smaller than the array FOV, a 64-point jiggle map produces a fully-sampled
map for both the LW and the SW arrays, by jiggling the secondary mirror to
fill in the undersampled arrays.  Finally, sources larger than the array FOV
are observed in scan-map mode.

The SCUBA Upgrade Project was designed to improve the sensitivity of
the instrument.  The first part of the project was completed in
October 1999, and included the installation of two wideband filters
centred at 450\,\micron{} and 850\,\micron{} (450W:850W).  The
wideband filters were designed to be more closely matched to the
atmospheric windows (\fig{filterfig}), and to be more sensitive than
their narrowband predecessors (450N:850N) under all weather
conditions.  The measured improvement is a few percent at
850\,\micron{} and a factor of 2 at 450\,\micron{}.  The improvement
at 850\,\micron{} is not due to the width of the filter, but to the
blocking filter that was installed to reduce contamination by infrared
light.  The overall spectral transmission characteristics of the
850-\micron{} waveband are driven by the edge filters and detector
feed-horn cut-off.  Thus, unfortunately, the measured response of the
850-\micron{} wideband filter is almost identical to that measured for
the narrowband filter.  For the purposes of this paper,
we will concentrate on the 450W:850W filters.

\section{Atmospheric attenuation}
\label{tau}

Determining the atmospheric attenuation of a source signal is critical for
calibrating submillimetre data.  Assuming a plane-parallel atmosphere:
\begin{equation}
I_m = I_{\circ}e^{- Tau A}
\end{equation}
where $I_m$ and $I_{\circ}$ are the signals incident at the telescope
and at the top of the atmosphere respectively, $A$ is the airmass (the
secant of the zenith distance), and Tau is the zenith sky opacity
\citep{jasonsecant}.  Thus, precise measurements of Tau must be taken 
frequently.  This is less crucial at 850\,\micron{}; in good weather,
\taulong{}$<$0.3, and at low airmass, ($<$1.5), a 20\% error in
\taulong{} alters the measured source flux density by 5-10\% at most.
However, in worse conditions and particularly at 450\,\micron{}, an
error in Tau can severely affect the measured source flux density.
For example, assuming a low airmass $<$1.5, a 20$\%$ error in
\taushort{} can alter the measured flux density by $\sim 50-80\%$.

Traditionally, Tau was derived by constructing a secant plot: the signal from
the source is measured as a function of airmass, and assuming the sky does not
change between measurements, the gradient of the plot gives Tau directly.
However, this method requires a meaningful number of measurements of a bright
source over a range of airmasses.  For any level of accuracy, the sky must
remain very stable over a long period of time.  \citet{jasonsecant}
demonstrate the difficulties of using this method to derive Tau.  We describe
here alternative methods that are able to track rapid variations in Tau.

\subsection{Skydip method}
\label{skydip}

SCUBA estimates the zenith sky opacity at the wavelength and azimuth of
observation by performing skydips.  Skydips measure the sky brightness
temperature as a function of elevation (usually between 80 and 15 degrees),
with absolute temperature calibration provided by hot and cold loads.  The hot
load is ambient temperature Eccosorb, and the cold load is a reflection of the
cold optics inside the cryostat, with an effective temperature of $\sim
60$\,K.  An aperture plane chopper unit, spinning at 2\,Hz, is used to switch
rapidly between the sky and the two loads.  The temperatures of the hot and
cold loads are measured and corrected for the emissivity and reflectivity of
the components in the optical path.  A model describing both the atmosphere
(assuming a plane-parallel form) and the optical system is then fit to the
data to calculate the zenith sky opacity:
\begin{equation}
J_{meas} = (1-\eta_{tel})\,J_{tel} + \eta_{tel}\,J_{atm} - bwf\,\eta_{tel}\,J_{atm}e^{-Tau A}
\end{equation}  
where $J_{meas}$ is the measured brightness temperature of the sky,
$\eta_{tel}$ is the transmission of the telescope, $J_{tel}$ is the
brightness temperature of a black-body radiating at the temperature of
the telescope, $J_{atm}$ is the brightness temperature of the
atmosphere, $bwf$ is the bandwidth factor of the filter being used
($1-bwf$ is the fraction of the filter bandwidth that is opaque due to
atmospheric absorption and, like Tau, is a function of water vapour
content), Tau is the zenith sky optical depth and $A$ is the
airmass of the measurement.  Technical details of this model are
presented in \appen{skydipmodel}, refer also to \citet{hazellthesis}.

\begin{figure}
\epsfig{file=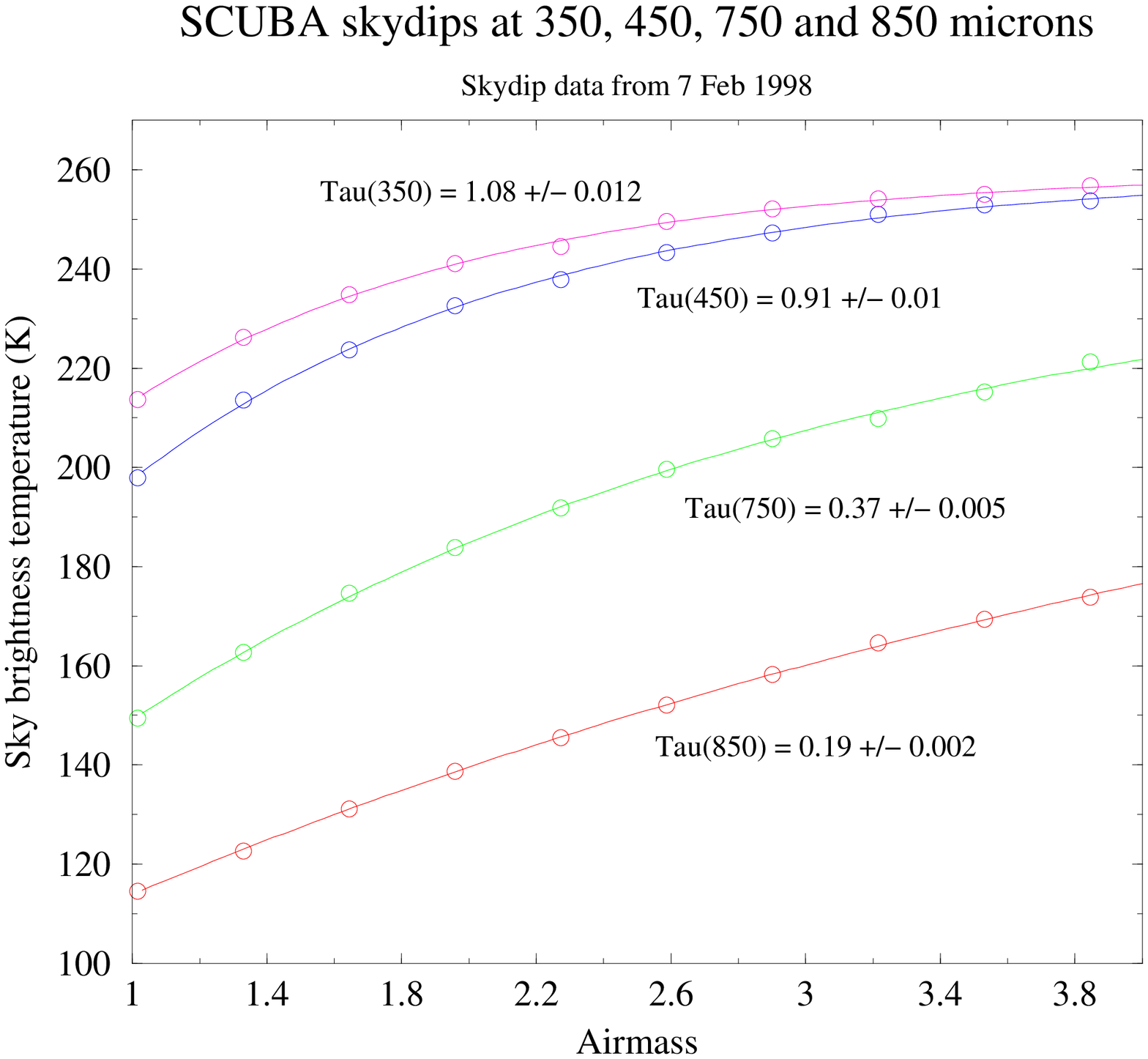,angle=-90,width=85mm}
\caption{Typical skydip examples at 350, 450, 750 and 850\,$\mu$m.  Open
  circles denote measurements taken with SCUBA; solid lines denote the model
  fits to the data.  The skydips shown here correspond to a precipitable water vapour of $\sim$\,0.7mm (see Equation 3).}
\label{skydipfig}
\end{figure} 

\fig{skydipfig} shows typical skydips taken with SCUBA at several
wavelengths.  In practice, the skydip method provides an accurate measurement
of Tau.  However, it takes $\sim 6$ minutes to perform a skydip.  Given this
overhead, it is only practical to perform a skydip every 1.5-2 hours, and
quite often the frequency is even less.  An on-the-fly skydip mode has
recently been commissioned, where data are taken continuously, reducing
the time taken to $\sim2-3$ minutes \citep{flydips}.

\subsection{CSO Tau Monitor}
\label{csofitsection}

\begin{figure}
\centering
\epsfig{file=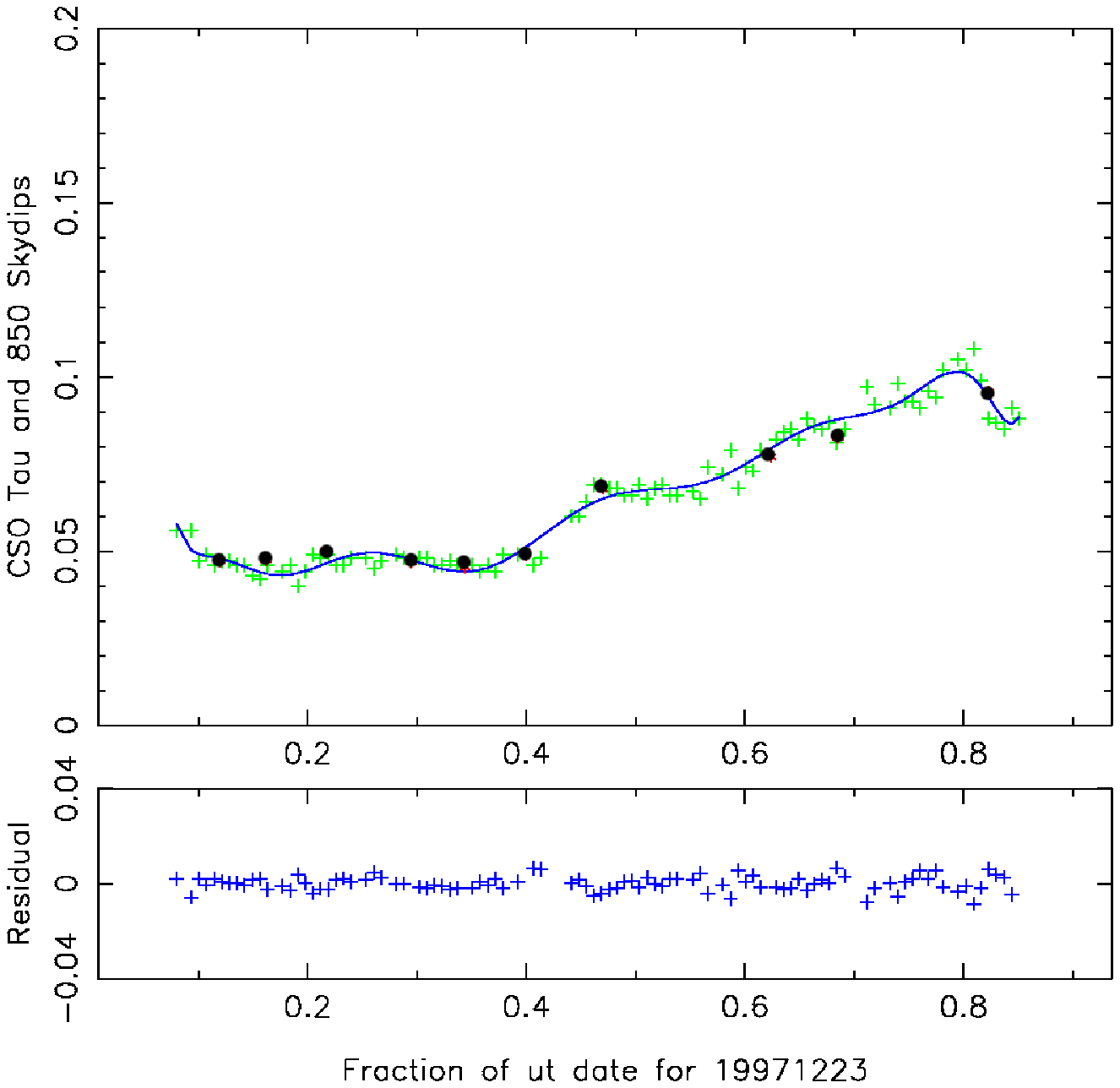,width=85mm}
\epsfig{file=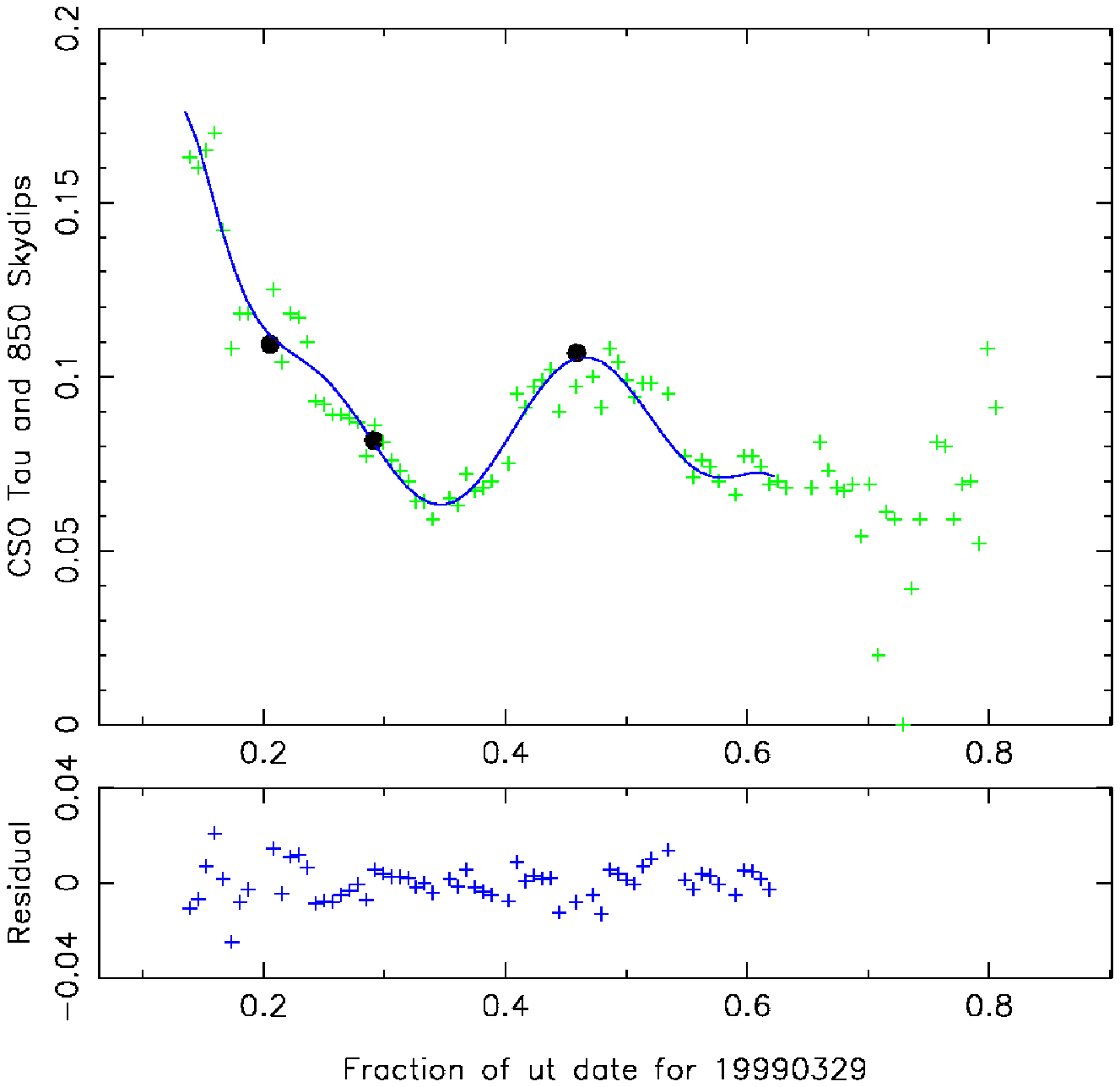,width=85mm}
\caption{CSO Tau data as a function of
  time (expressed as a fraction of the UT date) for two typical nights.  The
  CSO data are depicted by the `+' symbols.  The 850-micron{} skydips taken for
  each night, scaled to 225\,GHz values using the newly derived CSO Tau
  relations(\sec{taurelations}), are denoted by the solid circles.  The solid
  line is a polynomial fit to the CSO Tau data.  The fit residuals are also
  shown.}
\label{polyfits}
\end{figure}

The nearby Caltech Submillimetre Observatory (CSO) operates a 225\,GHz
(1.25\,mm) tipping radiometer, which performs a skydip every 10 minutes,
albeit at a fixed azimuth.  The precipitable water vapour of
the atmosphere is related to the CSO Tau as follows :
\begin{equation}
pwv = 20(Tau_{CSO} - 0.016)
\end{equation}
where the $pwv$ is in millimetres \citep{pwv}.

Comparing skydips taken with SCUBA to those taken at the CSO yields relations
between \taucso{} \& \taulong{} and \taucso{} \& \taushort{}.  If the scatter
about these relations is small, the CSO Tau monitor can be used to measure the
opacity more frequently than we can with SCUBA, with no additional overhead.

The CSO data show a significant amount of noise, but track long time-scale
($\sim$2 hour) variations very well. This noise is more than one would expect
given the measurement errors quoted in the CSO Tau archive.  It is probable
that this is due to instrumental noise and does not represent the behaviour of
the sky.  A polynomial can be fit to the data to track the large-scale
variations in CSO Tau as opposed to the (presumably) instrument noise.

In \Fig{polyfits}, example polynomial fits are presented for two
typical nights.  Using the relations derived in \sec{taurelations}, we
have also plotted the 225\,GHz Tau predicted by the 850-\micron{}
skydips taken each night.  It is striking how well the skydips track
the polynomial fits, even when one would imagine that the CSO Tau was
moving around too much to be useful.

Producing a composite picture of the night, using the CSO Tau, the SCUBA
skydips, and the polynomial fit, provides an additional level of quality
control.  It can give the observer a much better feel for how the atmosphere
was actually behaving on a given night, especially if conditions were
apparently unstable. Consider first the lower plot in \fig{polyfits}.  It is
clear that interpolating between the SCUBA skydips would give an erroneous
measurement of Tau.  In addition, the extreme scatter in CSO Tau indicates
that the end of the night was unusable.

The composite picture can provide further information about when data should
be treated with care.  For example, on some nights the CSO Tau and SCUBA
skydip data disagree with each other, on other nights both display unusually
high levels of scatter, indicating an inherently unstable night.  If, on the
other hand, the CSO Tau has a high-level of scatter but the SCUBA skydips
follow the polynomial fit, the scatter is unlikely to be representative of the
sky itself.

\subsection{Tau relations}
\label{taurelations}

\begin{figure*}
 \noindent
 \begin{minipage}[b]{.42\linewidth}
 \centering
 \epsfig{file=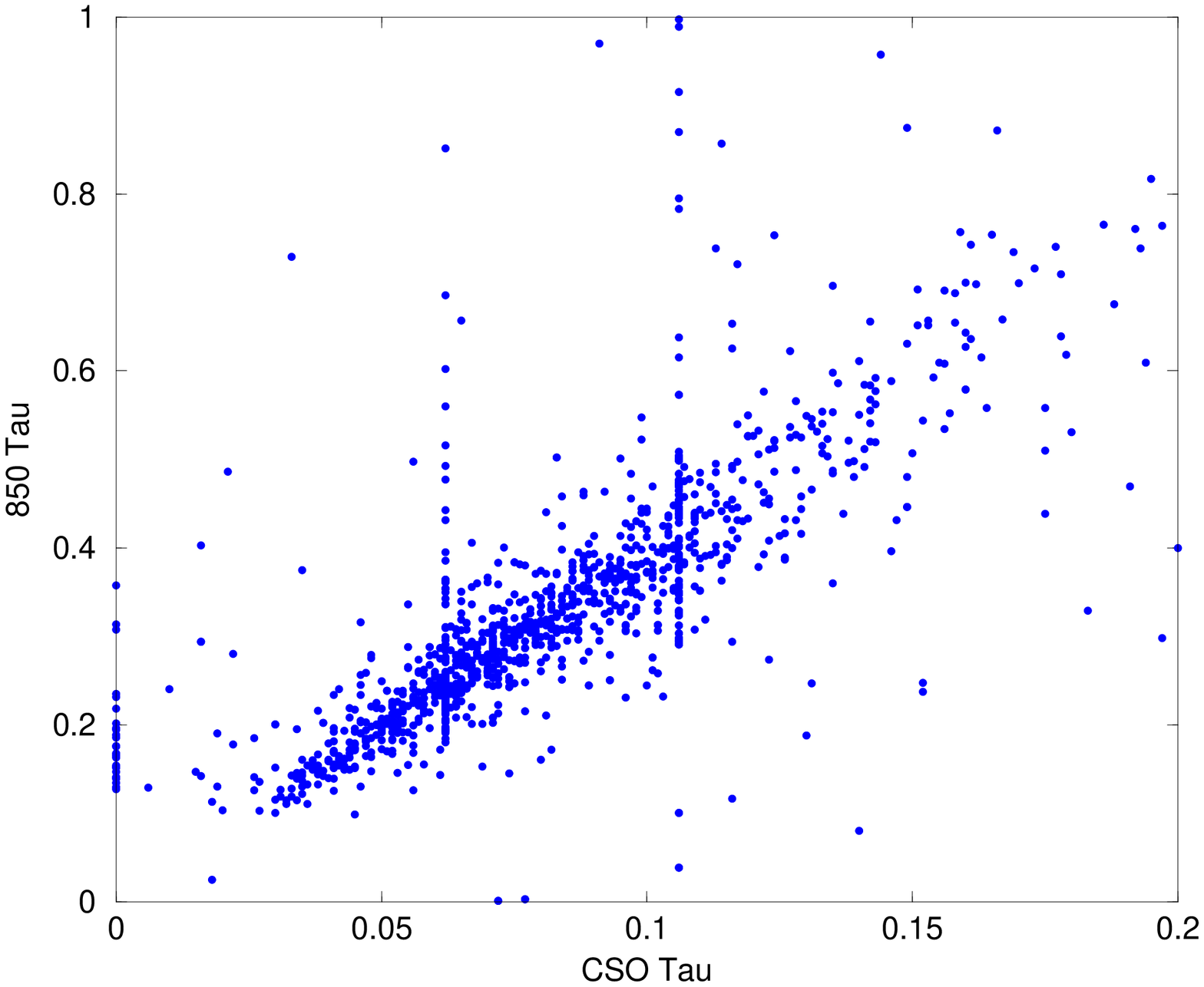,height=\linewidth,angle=270}
 \end{minipage}\hspace{17pt}
 \begin{minipage}[b]{.42\linewidth}
 \centering
 \epsfig{file=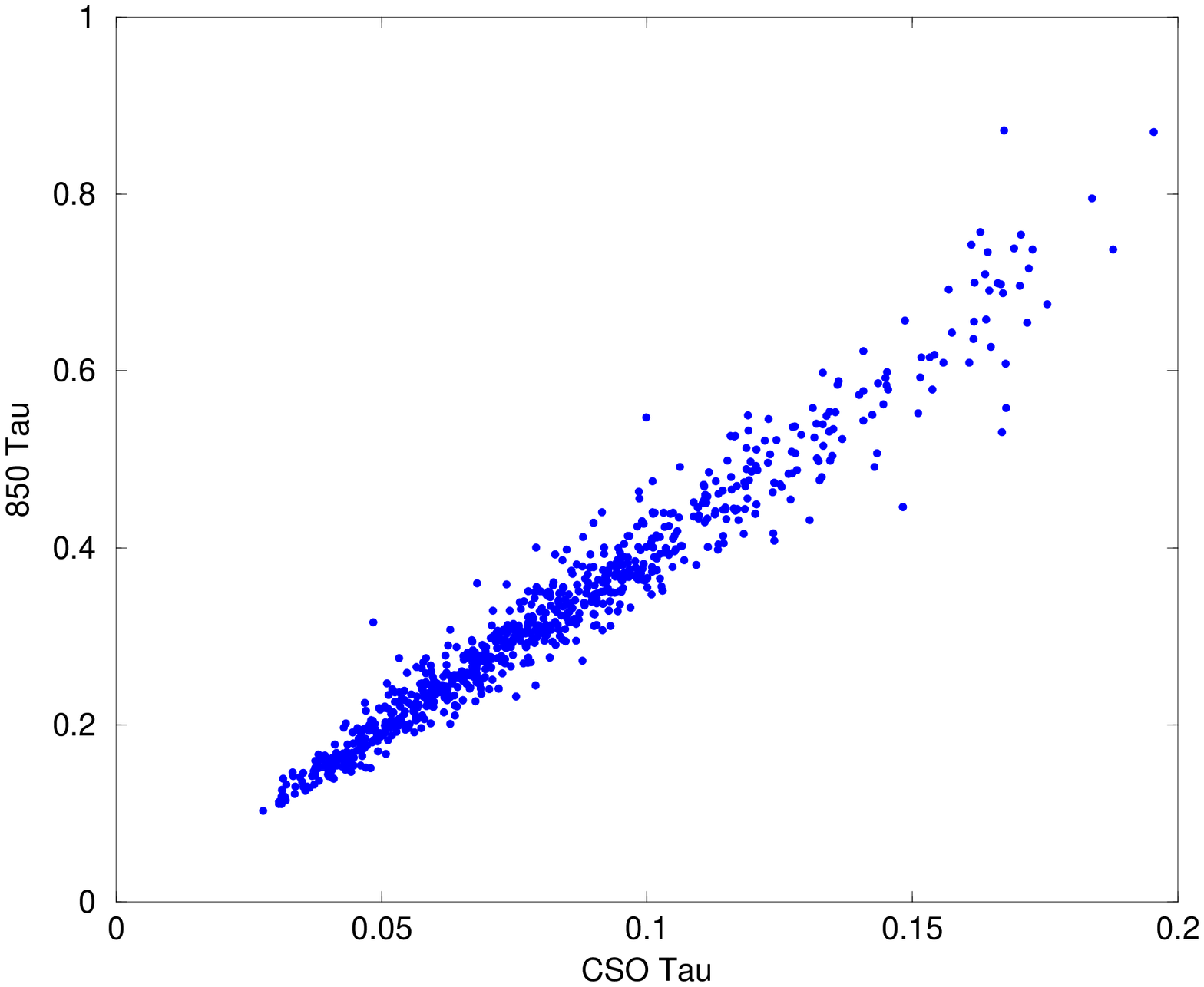,height=\linewidth,angle=270}
 \end{minipage}
\caption{The wideband \taulong{}-\taucso{} relation.  For the
  left plot, the CSO Tau data were taken directly from the archive, and every
  skydip observation was accepted.  The right plot shows the relation with
  \taucso{} calculated from the polynomial fits, and with the poor-fitting
  skydips having been removed.}
\label{reducedscatter}
\end{figure*}

\begin{table*}
\centering
\begin{tabular}{|ll||ccrr|}
\hline 
                &                       &\multicolumn{4}{|c|}{$Tau_{Y}=a(Tau_{X}-b)$}\\
Filter System   &Time Period            &$Tau_{Y}$      &$Tau_{X}$      &\multicolumn{1}{c}{$a$}                        &\multicolumn{1}{c|}{$b$}\\
\hline 
&\multirow{3}{4.6cm}{Feb. 04, 1998-Oct. 10, 1999 (pre-upgrade)}&\taulong{}        &\taucso{}      &3.99$\pm$0.02  &0.004$\pm$0.001\\
       450N:850N&                       &\taushort{}    &\taucso{}      &23.5$\pm$0.2 &0.012$\pm$0.001\\
                &                       &\taushort{}    &\taulong{}     &5.92$\pm$0.04  &0.032$\pm$0.002\\
&&&&&\\
&\multirow{3}{4.6cm}{Dec. 05, 1999-Sept. 30, 2000 (post-upgrade)}&\taulong{}      &\taucso{}      &4.02$\pm$0.03  &0.001$\pm$0.001\\
       450W:850W&                       &\taushort{}    &\taucso{}      &26.2$\pm$0.3 &0.014$\pm$0.001\\
                &                       &\taushort{}    &\taulong{}     &6.52$\pm$0.08  &0.049$\pm$0.004\\
\hline 
\end{tabular}
\caption{Tau relations for both the narrow and wide 450:850 filter systems, calculated using least-squares regression.  The corresponding errors are $1-\sigma$.  The relations have been constructed using the CSO polynomial fits and by ignoring poor-fitting skydips.}
\label{tabrelations}
\end{table*}

The time-resolution of the polynomial fits to the CSO Tau is
impressive, $\sim$ 2 minutes if the residuals are small, as they
typically are.  We can capitalise on this by deriving tight relations
between the CSO Tau and the opacity measured at SCUBA wavelengths.

These relations have been constructed taking the following points into account:

\begin{enumerate}
\item Skydips were discarded where the model failed owing to, for example, the
  atmosphere losing its plane-parallel nature or a cloud drifting overhead.
  In these cases the model returns unrealistic values for the fit parameters
  and/or unusually high fit residuals.  This is a more common occurrence at
  450\,\micron{}, where the shape and height of the atmospheric window are more
  sensitive to the presence of water vapour (see \fig{filterfig}).  We discarded $\sim$20$\%$ of the data at 850\,\micron{} and
  $\sim$50$\%$ of the data at 450\,\micron{}.  Note, we can still calibrate at
  450\,\micron{} when the skydip fit fails as we can extrapolate from either
  the CSO or the 850-\micron{} Tau.
\item We ignored data taken when the CSO Tau monitor was off-line.
\item If either the CSO Tau monitor or the skydip indicated a non-physical
  value of Tau, i.e. negative or zero, the observation was ignored.
\item The datasets were restricted to \taucso{}$<0.2$ (SCUBA is not used in
  weather conditions worse than this).
\item A model of the form $Tau_{Y}=a(Tau_{X}-b)$ was fit to the data to derive
  the Tau relations.
\end{enumerate}

\begin{figure}
\centering
\epsfig{file=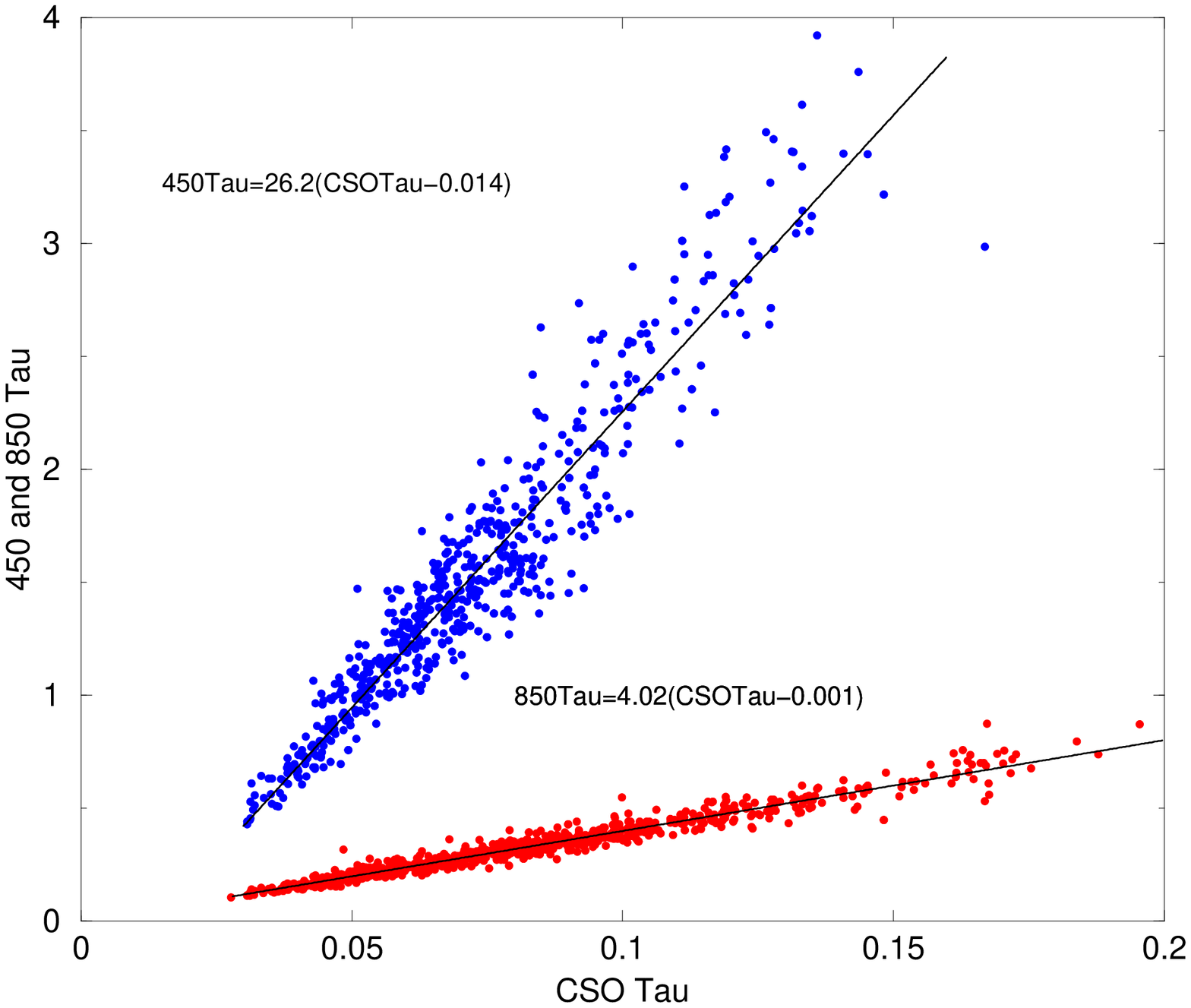,angle=-90,width=85mm}\vspace*{11pt}
\epsfig{file=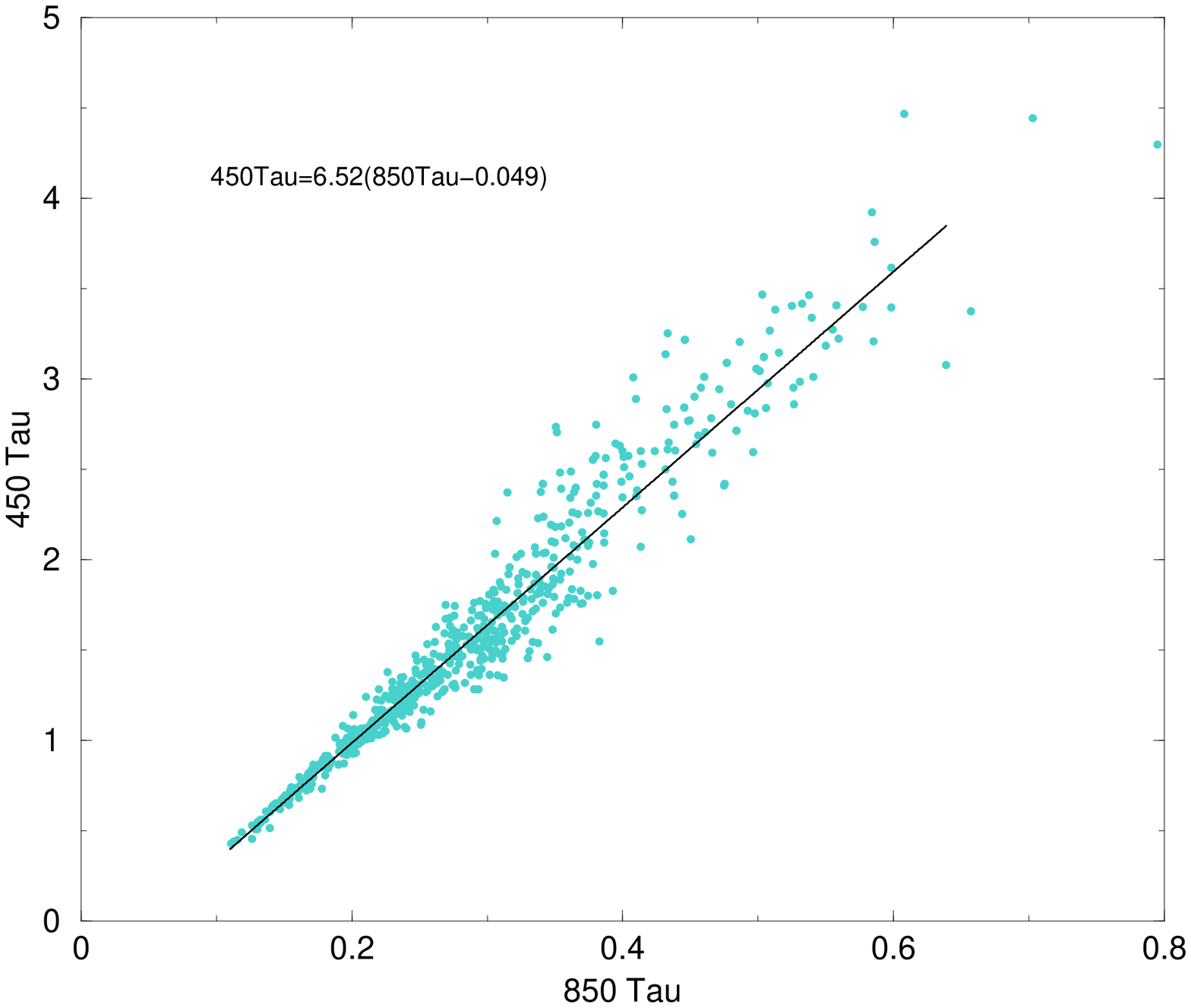,angle=-90,width=85mm}
\caption{Tau relations for the wideband 450W:850W filter system.  The top plot
  depicts the relationship between the SCUBA skydips and CSO Tau.  The bottom
  plot depicts the \taushort{}-\taulong{} correlation derived by comparing
  450-\micron{} and 850-\micron{} skydips.  Models of the form
  $Tau_{Y}=a(Tau_{X}-b)$ have been fit to the data in every case.  }
\label{widerelations}
\end{figure}

\fig{reducedscatter} shows how much the scatter in the relations has been
reduced by simply using the polynomial fits to estimate the CSO Tau and
ignoring untrustworthy skydips.  The final relations are presented in
\tab{tabrelations} and \fig{widerelations}.  They display relatively little
scatter, even in the poorest weather conditions, and can thus provide accurate
submillimetre calibration.

Comparing the wideband and narrowband filters, the wideband CSO Tau relations
are steeper at 450\,\micron{} but are almost identical at 850\,\micron{}.  At
first sight, the
difference at 450\,\micron{} is perhaps unexpected given the lower central
wavelength of the wideband filter (at 850\,\micron{}, the narrow and wideband
filters have almost identical central wavelengths).  However, the
450\,\micron{} wideband filter includes an H$_2$O line that is always present,
even in very dry weather conditions.  Thus, the wideband filter will always
have a steeper slope than the narrowband filter.

If the submillimetre opacity is due solely to water vapour, the
Tau relations for the different filters are expected to intercept the
origin: if there is no water vapour in the atmosphere, \taucso{},
\taulong{} and \taushort{} will all equal zero.  However, we have
found evidence for non-zero intercepts.  Assuming the straight-line
model can be extrapolated to the intercept, this could be
explained as follows: the relative contribution of ozone vs. water
vapour to absorption is greater at 225\,GHz than at 850\,\micron{} and
in turn greater at 850\,\micron{} than at 450\,\micron{}.  The ozone
contribution is relatively invariant for long periods of time, and
these offsets should be constant.

\section{Sky-noise}
\label{skynoise}

\subsection{Signatures of sky-noise}

\begin{figure}
\epsfig{file=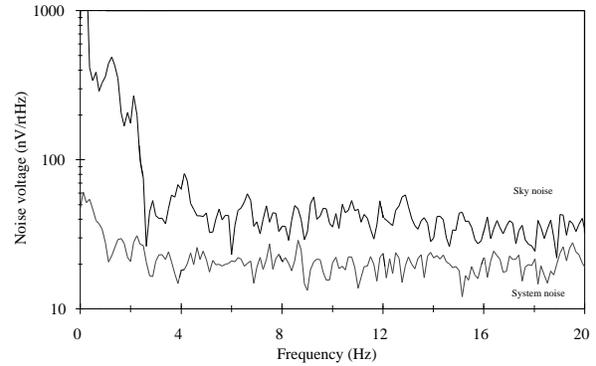,width=85mm}
\caption{Noise spectrum for LWA central pixel, showing 1/f noise tail of
  sky-noise.  The system noise trace is from the central bolometer looking at
  a 4\,K blank placed over the arrays.}
\label{noisespectrum}
\end{figure}

The atmosphere emits submillimetre radiation several orders of magnitude
larger than the signals we are trying to measure.  Variations in this emission
result in sky-noise, which manifests itself as a 1/f component in the noise
spectrum.  This is illustrated in \fig{noisespectrum}, where the 1/f component
extends out to 4\,Hz and is well above the system noise level.  Two-position
chopping at higher frequencies, $\sim 8$\,Hz, greatly reduces this sky-noise
contamination \citep{duncanchop}.

\begin{figure}
\epsfig{file=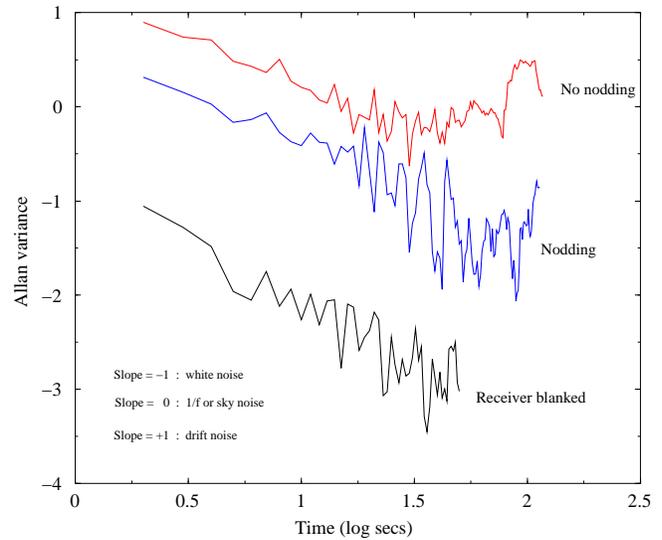,angle=-90,width=85mm}
\caption{Allan variance plot showing 1/f noise, white-noise and drift noise.} 
\label{allanvariance}
\end{figure}

As noted earlier, telescope nodding is also essential for the removal of
sky-noise.  This can be illustrated by the Allan variance, which is simply the
variance of the difference of two contiguous measurements \citep[e.g.,
][]{allan66,sk01}.  When plotted, a slope of $-1$ indicates white noise, a
slope of 0 denotes 1/f noise, and a slope of +1 is drift noise.
\fig{allanvariance} shows the Allan variance data as a function of time for
a typical observation with and without nodding (in both cases the telescope
was chopping at $\sim 8$\,Hz).  A clear 1/f component is present if the
telescope is not nodded: after around 20 seconds the noise stops
integrating down.  This component disappears when nodding is employed.

However, it is impractical to nod the telescope at the chop frequency.  Thus,
even with a chop/nod configuration, short-term temporal variations in the sky
emissivity, as well as spatial effects caused by chopping through slightly
different atmospheric paths, still exist.  This residual sky-noise must be
removed.  There are several ways to do this; filters can be designed to select
the most transparent parts of the atmospheric transmission window.  The
optical throughput should be single-moded, i.e. the minimum required to couple
to a point source with maximum spatial resolution and minimum background
\citep{duncanchop}.  Keeping the chop throw as small as possible, and in an
azimuthal direction (i.e.  chopping through the same atmospheric layer),
should also help.  In this paper we will concentrate on the benefits of having
a bolometer array to aid the removal of sky-noise.

\subsection{Removal of sky-noise}

In the submillimetre, the emissivity variations occur in atmospheric cells
that are larger than the array field-of-view \citep{remskypaper,borysnoise}.
Therefore, the noise should be correlated to a large extent across the array
and also between wavelengths.

One of the main advantages of SCUBA is that it contains an array of
bolometers.  When observing a compact source, the off-source bolometers can be
used to measure the sky signal on short timescales and hence remove any
residual sky-noise that chopping and nodding have failed to account for.

\begin{figure}
\epsfig{file=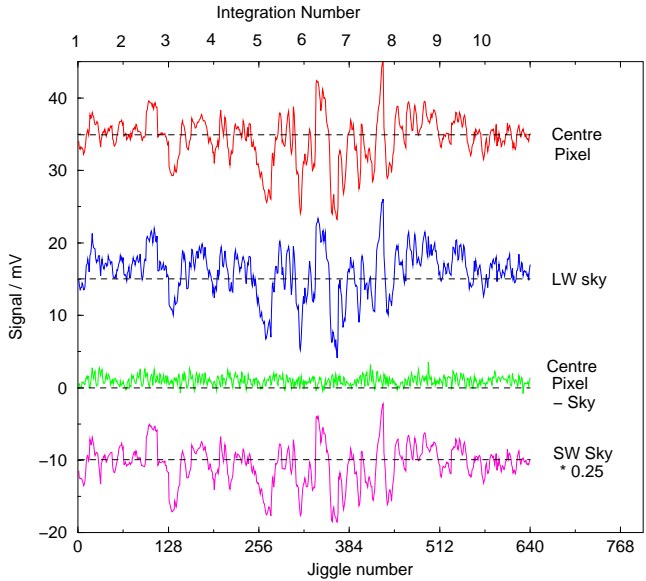,angle=-90,width=85mm}
\caption{Example of the identification and removal of sky-noise.  This figure 
first appeared in Holland et al. (1999) - note one integration comprises 128 seconds, split into 4 exposures of 16 jiggle points, with 1 second observed per jiggle point per nod position (there are two nod positions).}
\label{skysub}
\end{figure}

This is clearly illustrated in \fig{skysub}, where the signal of both
the centre on-source 850-\micron{} pixel and the sky (measured by
averaging the outer ring of bolometers) are shown.  The signals are
highly correlated, and subtracting the sky from the source reveals a
clear positive signal (in which the jiggle modulation is apparent).
Using this sky-subtraction method can improve the S/N significantly.
In \fig{skysub} we also display the 450-\micron{} centre pixel, and,
as expected, it is also highly correlated with the 850-\micron{} data.

\nocite{scubapaper}

\begin{figure}
\epsfig{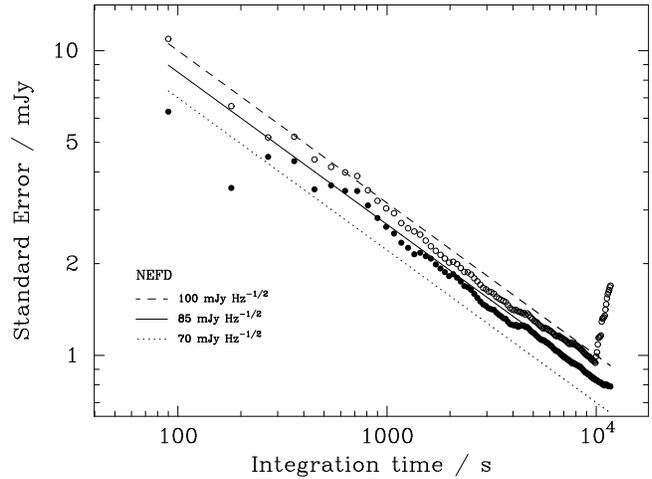}
\caption{Standard error evolution with time for a deep 850-\micron{} photometry
  observation.  For the open circles, only chopping and nodding have been used
  for sky cancellation.  For the solid circles, the sky-noise residual has
  been removed using the off-source bolometers.  This data was, for the most
  part, taken in uncommonly stable weather conditions.}
\label{noiseint}
\end{figure}

Failure to remove this residual sky-noise can, in the worst conditions,
degrade the sensitivity of the instrument by an order of magnitude.
\Fig{noiseint} displays the noise integrating down with time for a deep
850-\micron{} photometry observation.  Data are shown for the standard chop/nod
configuration, and also after the residual sky-noise has been removed using
the off-source bolometers.  We define the sensitivity, also known as the noise
equivalent flux density (NEFD), as the noise reached in 1 sec of integration.
It is clear that the NEFD of the instrument is significantly better if the
residual sky-noise is removed.  Furthermore, at the end of the observation,
there was a very obvious, large deterioration in the sky conditions.  This is
plainly visible in the case where the residual sky-noise was not removed.
Removing the residual takes this change of conditions into account, and allows
the noise to integrate down $\propto \sqrt{t}$.

For scan-map observations, removing the sky-noise is more difficult, as every
bolometer could be observing either the source or the sky at a given time.
The source can be removed from the datastream by the simple assumption that
the source structure is constant over time, while the sky is varying on
timescales of a few seconds.  The sky-emission noise can then be calculated
and removed from the data (Jessop et al. in preparation).

\subsection{Investigation of sky-noise}

For the purposes of this paper, we define sky-noise as the standard deviation
of the sky signal, with any errors associated with nodding removed.  The
dataset used here consists of deep photometry observations only, as it is easy
to specify the sky bolometers and the long observations provide good
statistics.

\begin{figure}
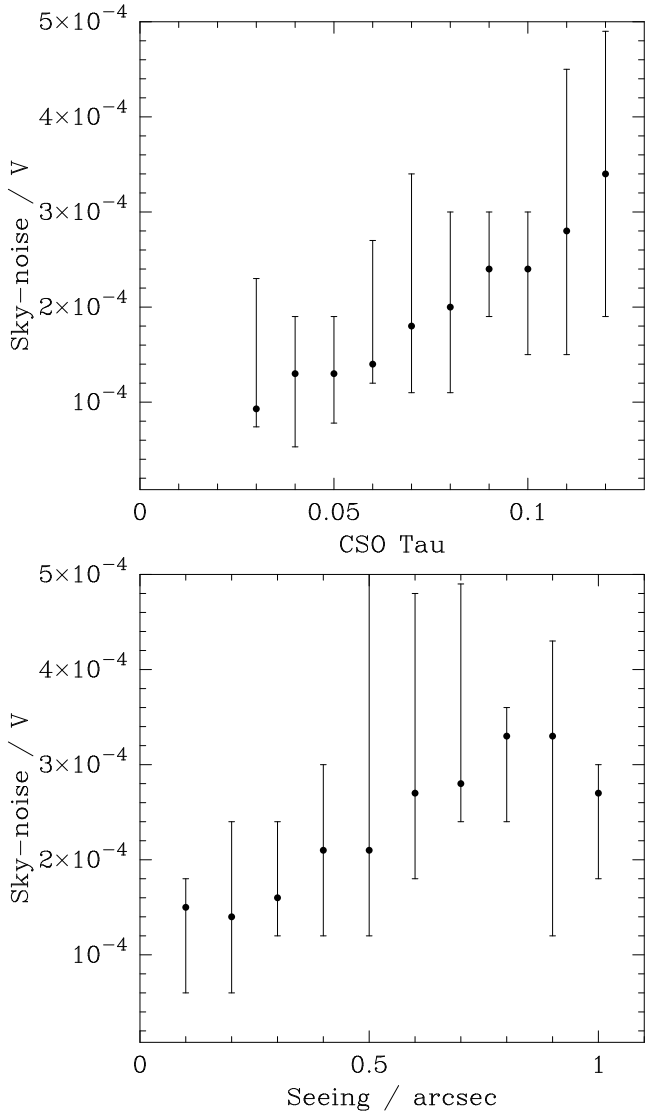

\epsfig{file=f10a.eps,angle=-90,width=85mm}
\epsfig{file=f10b.eps,angle=-90,width=85mm}
\caption{850-\micron{} sky-noise binned in CSO Tau (top plot) and seeing
  (bottom plot).  The data points are the mode of each bin, and the `errors'
  are the corresponding full-width half-maxima.}
\label{skynoisecorr}
\end{figure}

The relationship between sky-noise, seeing, and sky opacity is not
immediately obvious; the scatter in the data was too large to reveal
an obvious trend.  To overcome this, we binned the data.  This also
posed problems as the data were not Gaussian distributed and it was
not clear whether the mean (which is skewed by outliers) and
associated standard deviation (which is unnaturally large given the
scatter in the data) were useful quantities to measure.  Instead we
measured the mode and full-width half-maximum of the data in each bin.
These plots are shown in \fig{skynoisecorr}, and although the `errors'
are somewhat large, there is evidence for positive correlations
between sky-noise and seeing and between sky-noise and Tau.  There is,
however, no evidence to support the long-standing belief (from
single-pixel instruments) that sky-noise often increases in very dry
weather conditions.

\begin{figure}
\epsfig{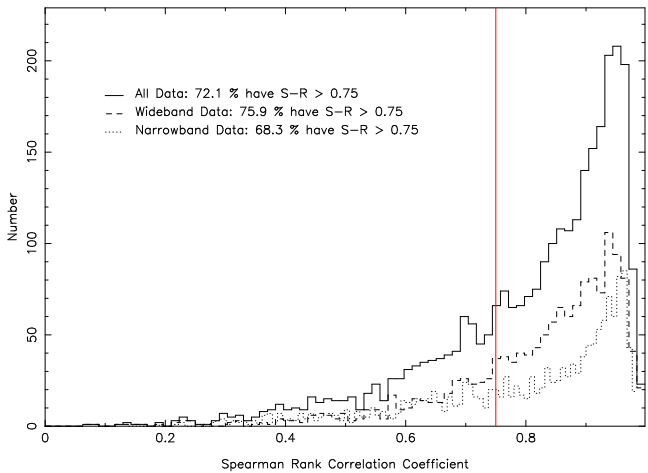}
\caption{Histogram of Spearman-rank correlation coefficients comparing
  850\,\micron{} and 450\,\micron{} sky signals.}
\label{spearhist}
\end{figure}

Considering the 850-\micron{} and 450-\micron{} sky signals for 20 minute
photometry observations, \fig{spearhist} presents a histogram of the
Spearman-rank correlation coefficients.  A value of +1 indicates a pure
positive correlation, 0 indicates no correlation, and -1 indicates a pure
negative correlation.  On the whole, the 850-\micron{} and 450-\micron{} data
appear to be highly correlated.  However, there are clearly times when the
strength of the correlation is low.

Given the known correlation between sky-noise at 850\,\micron{} and
450\,\micron{}, it has been suggested that the short wavelength array could be
used to provide sky cancellation at 850\,\micron{}.  This would be extremely
useful for faint sources where the 450\,\micron{} field-of-view is likely to
be source-free, but the 850\,\micron{} field is expected to contain several
sources or extended structure. 

\begin{figure}
\epsfig{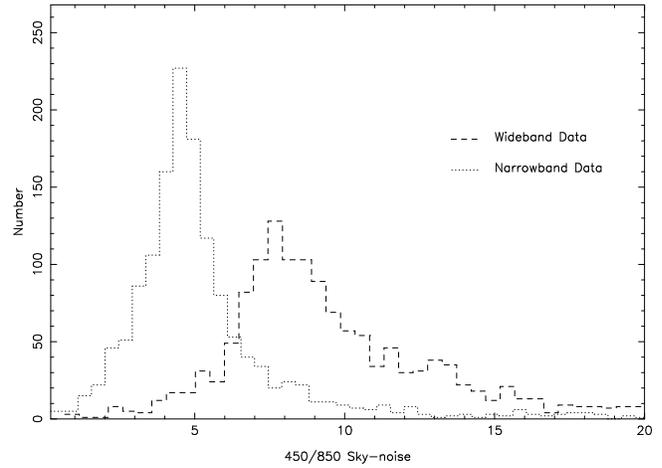}
\caption{Histogram of the ratio of 450-\micron{} sky signal to 850-\micron{}
  sky signal.}
\label{ratiohist}
\end{figure}

The ratio of the 450-\micron{} sky signal to the 850-\micron{} sky signal
(corrected for extinction) is displayed in \fig{ratiohist}.  For the
narrowband data, the median ratio is $\sim 4.5$ in agreement with
\citet{borysnoise} and \citet{remskypaper}; for wideband data, the median
ratio is $\sim 8$ and the distribution is wider.  The ratio shows less
dispersion if it is extinction corrected than if not, suggesting that the
water vapour creating the sky-noise is higher up in the atmosphere rather than
just above the telescope (where zero correction would be more suitable).

\begin{figure}
\epsfig{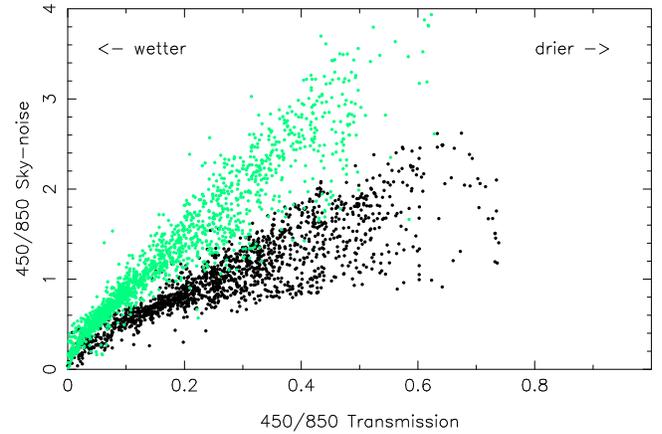}
\caption{The 450-\micron{}:850-\micron{} sky-noise (not corrected for
  extinction) ratio against the
  450-\micron{}:850-\micron{} transmission ratio.  The black circles denote
  the narrowband filter, the gray circles denote the wideband filter.}
\label{noisetrans}
\end{figure}

In \fig{noisetrans} we present the ratio of 450-\micron{}:850-\micron{}
sky-noise as a function of the corresponding transmission ratio.  Thus, for a
given transmission ratio, it is possible to use this plot to correct
850-\micron{} data for sky-noise using 450-\micron{} sky-noise data.  It is
clear that the sky-noise is larger for the wideband filter set than for the
narrowband.  This difference is to be expected as the 450-\micron{} wideband
filter is double the bandwidth of its narrowband predecessor, and contains an
${\rm H_2O}$ absorption line, while the two 850-\micron{} filters have almost
identical bandwidths (\nocite{duncan83}Duncan 1983 showed analytically that
for a sky-limited system $NEP \propto \Delta\nu$).  This excess sky-noise is
the cost of the improvement in sensitivity offered by increasing the
450-\micron{} bandwidth. However, as we have already demonstrated, this excess
is removable with a bolometer array.

A transmission ratio of 1 in \fig{noisetrans} corresponds to there being no
atmosphere between the telescope and the source of the sky-noise.  If the
sky-noise ratios in \fig{noisetrans} are extrapolated to a transmission ratio
of 1, we find values of $\sim 3.5$ and $\sim6.5$ for the narrowband and
wideband filters correspondingly.  These numbers are in good agreement with
the theoretical values derived in \appen{skynoiseappen}.  The
extinction-corrected ratios in \fig{ratiohist} are larger however,
indicating that we are over-correcting for extinction in calculating these
ratios.  This in turn suggests that the water vapour responsible for the
sky-noise lies somewhere in the middle-upper layers of the atmosphere, not at
the very top of the atmosphere itself.

\section{Submillimetre seeing}
\label{seeing}

\subsection{Measurement on Mauna Kea}
\label{seeingmsmt}

Submillimetre seeing arises from variations in the refractive index of the
atmosphere, principally due to the passage of water vapour through the beam.
This variation is measured on Mauna Kea by a phase monitor operated by
the Smithsonian Astrophysical Observatory (SAO). The SAO device comprises
two 1.8\,m dishes placed about 100\,m apart located near the JCMT. The dishes
point low in the eastern sky, and detect a signal at 12\,GHz from a
geostationary satellite. The difference in path lengths results in a phase
difference between the two received signals that changes slowly with the
oscillation of the satellite position. Turbulence in the atmosphere adds
noise to the phase difference. The phase difference is measured each
second and one minute's worth of data are analysed for the rms scatter to
provide the seeing at 12\,GHz over a 100\,m baseline.

The conversion from the rms of these phase fluctuations to an rms seeing 
for the 15\,m JCMT is suggested by \citet{masson91} as being:
\[{\rm JCMT~seeing~(arcsec~rms)  =  0.5 ~ SAO~seeing~(degrees~rms)}\]
Fifteen minute averages of these seeing data are available at the JCMT as part
of the Telescope Management System \citep{remo97}.  By the middle of the night
the JCMT seeing typically drops to 0.2\,arcsec, and observers consider seeing
$>1$\,arcsec to be `poor'.

\begin{figure}
\epsfig{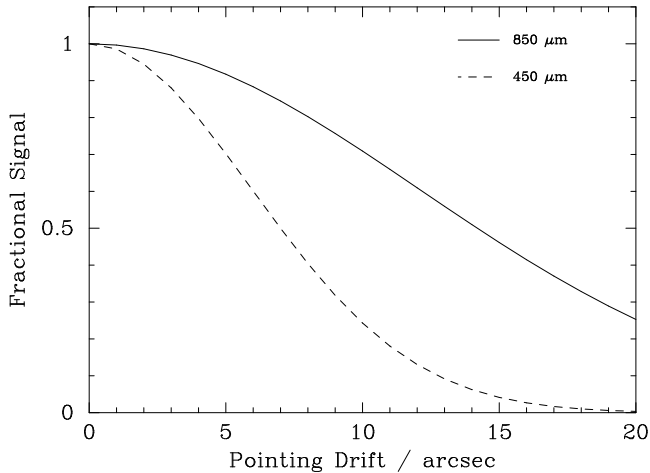}
\caption{Fractional signal as a function of pointing drift, assuming a FWHM of
  $14.2''$ at 850\,\micron{} and $7''$ at 450\,\micron{}.}
\label{pointingdrift}
\end{figure}

\subsection{Impact of seeing: broadening \& displacement}

If atmospheric turbulence on scales smaller than the telescope diameter is
significant then beam-broadening will occur. However, experiments by
\citet{churchhills90} indicate that these effects are small at the JCMT.  Beam
displacement results from large-scale, anomalous refraction, or
`tip/tilt' effects, and effective beam-broadening results if
observations take longer than the timescale of these fluctuations; the
time average of the beam motion being superimposed on the beam
profile.  Anomalous refraction timescales are of order 1\,s
\citep{olmi2001}, which is considerably shorter than standard JCMT
observations (minimum 18\,s), but comparable with the SAO monitor
rate.  JCMT jiggle maps (integration times $>32$\,s) therefore suffer
both minimal average pointing shifts and minimal image broadening as
the turblence tends to occur on scales larger than the telescope
diameter.

Additionally, analysis of JCMT tracking data shows that there is no
correlation between seeing and pointing excursions. Tracking data are long
jiggle maps of bright point sources, and each integration (32\,s worth) is
analysed for the location of the image centroid to yield information on
pointing offsets as a function of azimuth, for instance. Some tracking
datasets cover a sufficiently long time interval that significant changes in
seeing occur. Within such data there is no correlation between the change in
seeing and excursions in pointing.

Tracking data also tend to validate the conversion factor (0.5) between the
SAO phase (rms, degrees) and JCMT-seeing (rms, arcseconds), since the scatter
of the pointing offsets about some low-order polynomial fit is usually of the
same size as the JCMT-seeing.

\subsection{Do errors in computed refraction generate pointing errors?}

Relatively thin atmospheric layers (thicknesses $< 100$\,m, say) with
arbitrary values of temperature, pressure, and humidity have no impact upon
the overall refraction: the refraction towards the normal of a ray entering
such a layer is nullified by the refraction away from the normal upon exit.
Image displacement (a change in pointing) must therefore result from either
non-linear atmospheric structures or from uncompensated changes in local
humidity.

The latter hypothesis is simple enough to test using archived JCMT/SCUBA
pointing data, and no such relationship is found.  The formulae in
\appen{refacode} show that, to first order and at constant zenith distance,
elevation pointing corrections of $0.0768(h-20)$ arcseconds need to be applied
to account for the impact of local humidity, $h$(\%). In the absence of this
correction a plot of the change in elevation pointing against the change in
humidity would be expected to have a slope of 0.0768, whereas a plot of the
same data perfectly corrected for humidity will have a slope of zero. Analysis
of 3000 pairs of consecutive pointing measures yields a slope of
$-0.009\pm0.018$, commensurate with the latter scenario. The source of
residual pointing shifts therefore would seem to be excursions from the
idealized atmospheric model - i.e. turbulence - which is probably no great
surprise.

\subsection{Impact of pointing errors upon flux measurements}
\label{seeingflux}

Whatever their origin, pointing errors of size $\theta$ (arcseconds) cause a
signal of strength $S_o$ to be measured as
\begin{equation}
S  =  S_o~exp \left[ -4~ln(2)~\left(\frac{\theta}{FWHM}\right)^2\right]
\end{equation}
where FWHM is the full width at half-maximum of the Gaussian beam.
These are shown graphically in \fig{pointingdrift} in the cases of the JCMT operating
at 850\,\micron{} and 450\,\micron{}. 

As to the distribution of pointing errors as a result of refraction
noise, we have determined that the expected signal, $<S>$, is
related to the true peak signal, $S_o$, by
\[{\rm S_o/<S>~ = ~1~+~0.0275~\sigma^2~~for~a~14.2''~beam~at~850~\mu m}\]
\[{\rm S_o/<S>~ = ~1~+~0.1132~\sigma^2~~for~a~7''~beam~at~450~\mu m}\]
where $\sigma$ is the rms refraction noise.  The full derivation of these
equations can be found in \appen{seeingmaths}.

\section{The effect of the atmosphere on sensitivity}
\label{sensitivity}

We have recently made a concerted effort to characterise SCUBA's performance.
One aspect of this work was the creation of a database of calibration
observations.  The key result from studying this data is that the flux
conversion factors (FCFs) are quite stable, to an accuracy of 5\% at
850\,\micron{} and 20\% at 450\,\micron{} \citep{timfcfpaper}.  Using this
database, we have created a homogeneous dataset of $\sim 4000$ calibrated
photometry observations, to better understand the effect of the atmosphere on
sensitivity.

\begin{figure}
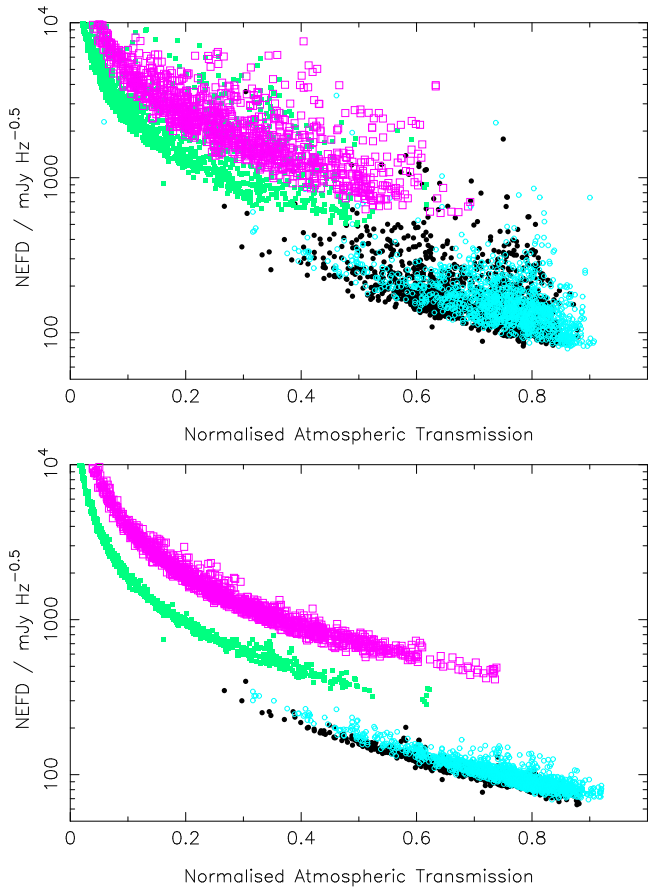

\epsfig{file=f15a.eps,angle=-90,width=85mm}
\epsfig{file=f15b.eps,angle=-90,width=85mm}
\caption{SCUBA sensitivity as a function of normalised sky transmission for
  the 850W (black circles), 850N (open blue circles), 450W (green squares),
  and 450N (open pink squares) filters.  For the top plot, only chopping and
  nodding were used for sky cancellation.  For the bottom plot, the off-source
  bolometers were also used for sky cancellation.}
\label{nefdtrans}
\end{figure}

\fig{nefdtrans} presents the NEFD as a function of normalised sky transmission
for a chop/nod configuration, and also using the off-source bolometers
to remove the residual sky-noise.  The NEFD is significantly lower and
is inherently more stable if the off-source bolometers are used for
sky-cancellation.  Data are shown both for the wideband filters, and
the pre-upgrade narrowband filters.  It is worth noting that the
450-\micron{} wideband filter (which is better matched to the entire
atmospheric window but includes an H$_2$O line) is considerably more
sensitive under all conditions.  There is little difference between
the narrow and wideband 850-\micron{} filters, which is to be expected
because, as noted in \sec{scuba}, the overall filter transmission
profiles are almost identical as measured in SCUBA.

\begin{figure}
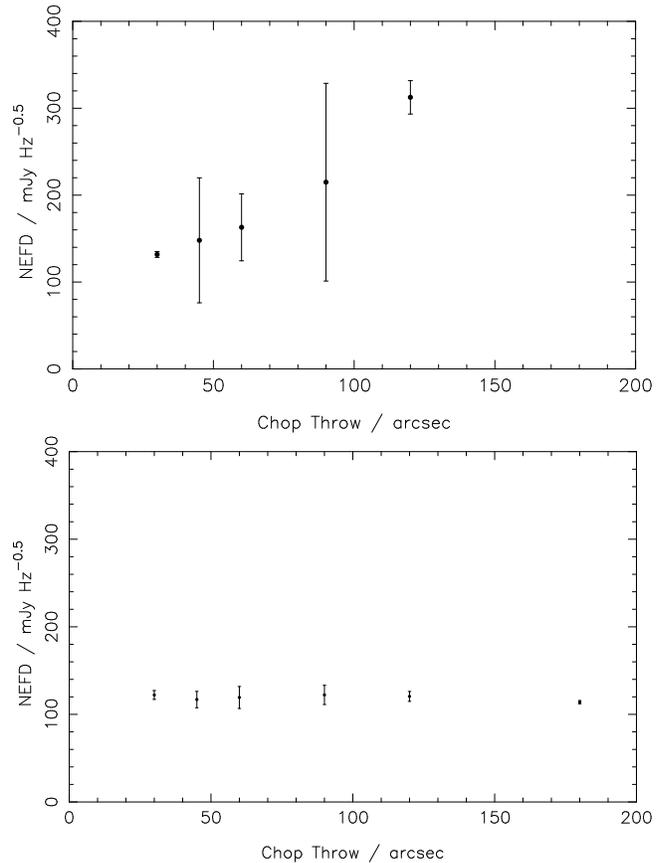

\epsfig{file=f16a.eps, angle=-90, width=8.5cm}
\epsfig{file=f16b.eps, angle=-90, width=8.5cm}
\caption{NEFD vs. chop throw for both the wide and narrowband 850-\micron{}
  filters.  The dataset is restricted to the normalised sky transmission lying
  between 0.5 and 0.75, and the values are normalised to a transmission value
  of 0.65.  The top plot is for a chop/nod method of sky cancellation and is
  restricted to observations with an azimuthal chop throw.  In the
  bottom plot off-source bolometers are used to remove the residual sky-noise,
  and there is no restriction on the chop throw angle.}
\label{nefdthr}
\end{figure}

We have also considered how the sensitivity of the instrument is affected by
the chop throw.  \fig{nefdthr} presents the data for two cases: a standard
chop/nod method of sky cancellation, and a chop/nod method where the
off-source bolometers are used to remove residual sky noise.  Although the
error bars are somewhat large, the standard chop/nod method seems to support
the long-held belief that sensitivity decreases with chop throw (e.g. Church
et al. 1993; Duncan et al. 1995).  However, when using the array and when
off-source bolometers are used for sky cancellation, large chop throws {\em do
  not} degrade the sensitivity.  In this case, the NEFD vs. chop throw plot is
perfectly flat out to a chop throw of 180 arcsec.  The only variation is due
to an increase of the FCF with chop throw, of the order of 10\% between 45 and
120 arcsec \citep{timfcfpaper}.

\section{Summary and the future}
\label{future}

As described in this paper, we have a good grasp of the atmospheric limitations
in the submillimetre and how best to overcome them.  Our main findings can be
summarized as follows:

\begin{enumerate}
\item Tau relations between the CSO Tau at 225\,GHz and both the
  850\,\micron{} and the 450\,\micron{} SCUBA filters.  The relations display
  relatively little scatter.
\item For accurate sky cancellation, it is essential to use off-source
  bolometers to monitor and remove the sky-noise.
\item There is evidence for positive correlations between sky-noise and
  seeing, and sky-noise and sky opacity,but the correlation coefficient varies
  significantly depending on the dataset \fig{spearhist}.
\item 850-\micron{} and 450-\micron{} sky-noise are clearly correlated, but
  there are times when the correlation is low.
\item The JCMT beam is not significantly broadened by seeing.
\item There is no obvious correlation between seeing and pointing excursions.
\item If off-source bolometers are used for sky cancellation, chopping as
  far as 180 arcsec (in any direction) does not affect the sensitivity of
  the instrument.
\end{enumerate}

In the future, as our understanding grows, as technological advances are made,
and as the next generation of submillimetre telescopes are destined to higher
and drier sites, we will be able to do even better.  Some things to look
forward to are:

\begin{enumerate}
\item Line-of-sight radiometers which give real time estimates of Tau in the
  direction you are observing.  One of these radiometers is now in operation
  at the JCMT (e.g. Wiedner et al. 2001\nocite{martina2001}, Phillips et al.
  in preparation).
\item Instantaneous estimates of the line-of-sight opacity should be possible
  using a hot load, cold load, and sky observation at the elevation of the
  source, given a database of sky temperature versus elevation information
  (e.g. Smith, Naylor \& Feldman 2001 \nocite{irma}).
\item High-speed data sampling leading to a more efficient observing mode
  called `DREAM' \citep{dreammode}.  In DREAM mode, the functions of chopping
  and jiggling the secondary mirror are combined to a single step action,
  eliminating the need to sample empty sky for half the time. Another gain
  from high-speed sampling is real-time suppression of sky-noise.
\item Adaptive optics for the submillimetre.  The Large Millimetre Telescope
  (LMT) is currently designing a radiometric wave-front sensor which will
  measure the tilt of the incoming wavefront to compensate for
  atmosphere-induced pointing errors (For further details refer to the LMT
  website: http://www-lmt.phast.umass.edu/).
\item DC-coupled fully sampled arrays such as SCUBA 2 \citep{scuba2holland,
    scuba2robson}, which would remove the necessity to chop and nod for
  sky-noise removal.  This is an advantage: there is no chance of chopping
  onto a nearby source, and there will be a sensitivity gain as all the time
  is spent looking at the source instead of half the time looking at the sky.
  In addition, chopping limits the size-scales we see in maps to be no more
  than a few times the chop throw.  Finally, image reconstruction techniques
  for two-beam chopping tend to propagate noise, and so with only a
  single-beam on the sky this problem will be minimsed.
\item In the future, submillimetre telescopes will be built at sites where the
  atmosphere is largely transparent, for example Chajnantor in Chile at 17,000
  ft. (ALMA) and the South Pole.  The extreme cold at the South Pole results
  in very small amounts of water vapour in the atmosphere, and thus sky-noise
  is much lower than other sites \citep{southpole}.
\item Interferometers such as ALMA have a unique advantage: the
  atmospheric signals decorrelate, and therefore sky-noise should not be a problem.
\end{enumerate}

\section*{ACKNOWLEDGMENTS}

We wish to thank the Canadian co-op students who have contributed to this
work: Ed Chapin, Jeff Wagg, and Karl Kappler.  We also wish to thank David
Naylor and his group for the SCUBA FTS measurements of the filter profiles.
This work has made use of the Tau archive maintained by the Caltech
Submillimetre Observatory, and the phase monitor operated by the Smithsonian
Astrophysical Observatory.  We acknowledge the support software provided by
the Starlink Project which is run by CCLRC on behalf of PPARC.  The JCMT is
operated by the Joint Astronomy Centre, on behalf of the U.K.  Particle
Physics and Astronomy Research Council, the Netherlands Organisation for Pure
Research, and the National Research Council of Canada. Further details and
updated information can be found on the SCUBA World-Wide Webpage at URL:
http://www.jach.hawaii.edu/JCMT/Continuum$\_$observing/\\continuum$\_$observing.html.

\appendix
\section{Skydip Model}
\label{skydipmodel}

When performing a skydip, SCUBA measures the sky brightness temperature as a
function of airmass.   We present a multi-layer model of the atmosphere
\citep{hazellthesis} which, when compared with the data, yields the zenith sky
opacity.  The model at each wavelength takes the form:
\begin{equation}
J_{meas} = (1-\eta_{tel})\,J_{tel} + \eta_{tel}\,J_{atm} - bwf\,\eta_{tel}\,J_{atm}e^{-Tau A}
\end{equation}  
where $J_{meas}$ is the measured brightness temperature of the sky,
$\eta_{tel}$ is the transmission of the telescope, $J_{tel}$ is the
brightness temperature of a black-body radiating at the temperature of the
telescope, $J_{atm}$ is the brightness temperature of the atmosphere, $bwf$
is the bandwidth factor of the filter being used ($1-bwf$ is the fraction
of the filter bandwidth that is opaque due to atmospheric absorption and,
like Tau, is a function of water vapour content), Tau is the
zenith sky optical depth and $A$ is the airmass of the measurement.

Of these parameters, $J_{meas}$, $J_{tel}$ and $A$ are
known. $J_{atm}$ can be estimated from the ambient air temperature
at ground level using a model for the behaviour of the observing layer
above the telescope, as described below. $\eta_{tel}$ may be fitted
to the data for every skydip and, because it does not vary with
atmospheric conditions, a reliable `average' value can be derived from
many observations. Thus, there are two remaining free parameters, $\tau$
and $bwf$, that must be derived from the fit.

$J_{atm}$ is calculated from $T_{amb}$, the ambient air temperature, by
assuming that the sky emission is dominated by a single absorber/emitter
whose density falls exponentially and temperature linearly with height. In
this case it can be shown that:
\begin{eqnarray}
J_{atm}\,=&J_{amb} \int_{0}^{40} Ak\,exp\left(-\frac{h}{h_2}\right) \times \nonumber \\
&exp\left[Akh_2\left(\exp\left(-\frac{h}{h_2}\right)-1\right)\right]
\left(1-\frac{h}{h_1}\right)\,dh
\end{eqnarray} 
where $h_1$ is $J_{amb}/6.5$ to give a 6.5\,K fall in temperature per km
height, $h_2$ is the scale height of the absorbers (2\,km), $A$ is the
airmass and $k$ the extinction per km.

If we approximate the result of the integral by: 
\begin{equation}
J_{atm} = J_{amb} X_{g} \left[1-\exp\left(-A k
h_2\right)\right]
\end{equation}
it can be shown that $X_{g}$ has the form:
\begin{equation}
X_{g} = 1 + \frac{h_2
T_{lapse}}{T_{amb}}\exp\left(-\frac{A
Tau}{X_{gconst}}\right)
\end{equation}
where $T_{lapse}$ is the temperature drop per kilometre altitude
($-6.5\,$K/km) and $X_{gconst}$ is a constant determined empirically and
has a value of 3.669383.

\section{Theoretical derivation of sky-noise ratio}
\label{skynoiseappen}

 If we consider two lines of sight separated by an angle theta and
 identical to each other except for the fact that one line of sight has a
 cloud of water vapour at a distance to which the line of sight opacity is
 $\tau_{cloud}$, and the cloud itself has optical depth $\delta\tau$, then
 the difference in surface brightness between the two lines of sight
 (measured in units of brightness temperature) is simply

 \begin{equation}
 \delta J = exp(-\tau_{cloud}) ~\delta\tau~(J_{cloud} -J_{inc})
 \end{equation}

 where $J_{cloud}$ is the physical temperature of the cloud, and $J_{inc}$
 is the incident flux on the cloud from the upper atmosphere. In all cases
 it is likely that $J_{cloud}$ is greater than $J_{inc}$ so that the
 observed effect of the cloud is to increase the surface brightness along
 that line of sight.

 If the cloud is close to the top of the water vapour column density then
 $J_{inc}$ is negligible compared to $J_{cloud}$ and one can simply write:
 \begin{equation} \delta J = exp(-\tau_{cloud}) ~\delta\tau ~J_{cloud}. 
 \end{equation}

 If one considers the signal at two wavelengths $\lambda1$, $\lambda2$ then
 the ratio is
 \begin{equation}
 \frac{\delta J_{\lambda1}}{\delta J_{\lambda2}} =
 \frac{exp(-\tau_{\lambda1})}{exp(-\tau_{\lambda2})} 
 \frac{\delta\tau_{\lambda1}}{\delta\tau_{\lambda2}},
 \end{equation}
 where we have removed the subscripts cloud.
 One can express this in terms of power by simply invoking the
 Rayleigh-Jeans law $B(\lambda) \propto J/(\lambda^2)$ so that,
 \begin{equation}
 \frac{\delta B_{\lambda1}}{\delta B_{\lambda2}}=
 \frac{exp(-\tau_{\lambda1})}{exp(-\tau_{\lambda2})}
 \frac{\delta\tau_{\lambda1}}{\delta\tau_{\lambda2}}
 \frac{\lambda1^2}{\lambda2^2},
 \end{equation}
 and also to measured volts by applying an appropriate FCF
 \begin{equation}
 \frac{\delta V_{\lambda1}}{\delta V_{\lambda2}}=
 \frac{exp(-\tau_{\lambda1})}{exp(-\tau_{\lambda2})}
 \frac{\delta\tau_{\lambda1}}{\delta\tau_{\lambda2}}
 \frac{\lambda1^2}{\lambda2^2}\frac{ FCF_{\lambda2}}{FCF_{\lambda1}}.
 \end{equation}
 
 In the case where the opacities are simply proportional to water vapour
 column density (as we have shown in this paper) then
 $\delta\tau_{\lambda1}/\delta\tau_{\lambda2}$ is a
 constant, and we expect the ratios of the signal at the two wavelengths to
 be proportional to the ratios of transmission from the cloud to the
 telescope at the two wavelengths. One also expects
 \begin{equation}
 \frac{\delta J_{\lambda1}}{exp(-\tau_{\lambda1})} \propto
 \frac{\delta J_{\lambda2}}{exp(-\tau_{\lambda2})}
 \end{equation}
 where one has to be careful to note that the value of $\tau$ is to the
 cloud, not all the way to the top of the atmosphere.
 
 For the SCUBA filters the ratios of $\delta\tau_{850n}/\delta\tau_{450n}$ and
 $\delta\tau_{850w}/\delta\tau_{450w}$ have been measured earlier in this
 paper. \citet{timfcfpaper} have measured the FCFs from point sources using a
 60\,arcsecond radius aperture, which are also (by symmetry) valid for a
 uniform extended source of size 60\,arcseconds. In the section on sky noise
 in this paper most of the photometry observations were taken with
 60\,arcsecond chop throws so that the skynoise measured is a measure of the
 difference in signal between two largely (but not exactly) overlapping
 regions of diameter 15\,meters, 60\,arcseconds apart. If the cloud is at a
 height, h above the telescope then the diameter of the beam in steradians is
 15/h(m) or 25\,arcminutes at a height of 2\,km. This explains why the sky
 noise is uniform over the array.

 In Fourier space we can consider the beam as a high spatial frequency
 filter of the form
\begin{equation} exp\left(\frac{-k^2}{15}\right)
\end{equation}
 i.e. one that filters out spatial scales $<$ 15\,m
 and the chop as a low spatial frequency filter of the form
\begin{equation}sin(2~\pi~k~c~h) \end{equation}
where $k$ is the wavenumber, $c$ is the chop throw in steradians, and $h$ is
the height of the cloud in metres.  This can be simplified to $2\pi kch$
for the frequencies that are passed by the beam.  The effect of having two
differing spatial frequency filters is to couple only weakly (at a level
$c h / 15$) to clouds of approximate spatial scale 15\,m.  Clouds
smaller than this are filtered out by the beam, clouds larger than this are
filtered out by the chop throw. For typical values ($h=2$\,km,
$c=60$\,arcseconds), the coupling is only 4 percent - illustrating how
effective chopping is as a first stage to removing sky noise.

 Although strictly speaking the FCFs
 calculated by \citet{timfcfpaper} are only valid for extended far field
 objects, one can use the ratio of the values measured as an estimate of
 the ratio of the FCFs which would be valid for nearfield extended objects
 (i.e. the cloud in question). The validity of this assumption can
 then be assessed by comparison with the data.

 If one therefore uses the various values discussed in the previous
 paragraph then one finds:

 \begin{equation}
 \frac{\delta V_{450W}}{\delta V_{850W}}=6.64~\frac{exp(-\tau_{450W})}{exp(-\tau_{850W})},   
 \end{equation}
 \begin{equation}
 \frac{\delta V_{450N}}{\delta V_{850N}}=3.32~\frac{exp(-\tau_{450N})}{exp(-\tau_{850N})}
 \end{equation}
 and
 \begin{equation}
 \frac{\delta B_{450W}}{\delta B_{850W}}=23.2~\frac{exp(-\tau_{450W})}{exp(-\tau_{850W})},   
 \end{equation}
 \begin{equation}
 \frac{\delta B_{450N}}{\delta B_{850N}}=21.2~\frac{exp(-\tau_{450N})}{exp(-\tau_{850N})}   
 \end{equation}
 Given that we have shown we are coupling to sources larger in scale than
 60\,arcseconds, we could further correct these values by the ratio
 \begin{equation}
  \frac{\eta_{450(cloud)}}{\eta_{850(cloud)}}~\frac{\eta_{850(60")}}{\eta_{450(60")}}
\end{equation}
where $\eta$ is the coupling.  However, given uncertainties in the cloud's
angular size (principally due to uncertainties in its height), and the fact
that the ratio of the coupling values is likely to vary less significantly
than the individual values, we choose not to.

\section{Refraction}

\subsection{Refraction Code at JCMT}
\label{refacode}

The primary purpose of the current JCMT refraction model is to provide
coarse corrections to pointing based on local atmospheric conditions. It
assumes a well-behaved, non-turbulent atmosphere.  To first order,
refraction displaces images towards the zenith by an amount
\begin{equation}     R  =  A * tan(z) \end{equation}     
where $z$ is the zenith distance, and $A$ is a function of
the local atmospheric parameters :
\begin{enumerate}
\item $T$ - Temperature (above some mean, chosen as 4\,K for Mauna Kea)
\item $p$ - \% Pressure change from the MK standard of 624\,mb
\item $h$ - \% Humidity
\end{enumerate}

Grids of atmospheric models were generated and refraction calculated
at several zenith distances by integrating through the atmosphere. 
Simple functional forms of the results were sought that would enable
calculation of refraction from local atmospheric conditions. The optical-
and millimetre- $A$-terms in the formula above take principally these forms
(units are arcseconds) :
 
  \begin{equation}A   = 35.893 - 0.00067(h-20) - 0.135(T-4) + 0.371p\end{equation}
at 0.55\,\micron{} and:
  \begin{equation}A   = 36.800 + 0.0768(h-20) - 0.0294(T-4) + 0.371p\end{equation} 
at 1\,mm.

- i.e. optical refraction is dependent mostly upon temperature, while 
submillimetre refraction is dependent mostly upon humidity. The 1\,mm
formulation used at JCMT contains additional, less significant 
terms. Its values were compared to the integration results throughout
the grid and were found to be accurate to better than 1\,arcsec for zenith
distances less than 80\,degrees under all but the most extreme conditions.

\subsection{Expectation Values of Signal Levels in the Presence of Refraction}
\label{seeingmaths}

This appendix describes the use of the pointing rms values determined from
the SAO phase monitor to give an indication of the expectation values of
the signal levels, relative to the values for no refraction 'noise'.
The assumptions are
\begin{enumerate}
\item the refraction noise can be described by a Gaussian process, which
      in two dimensions gives the Rayleigh distribution:
  \begin{equation}
  P_{\theta} = \frac{\theta}{\sigma^2}\,exp\left(\frac{-\theta^2}{2\sigma^2}\right)
  \label{eq:rayleigh}
  \end{equation}
  where P$_{\theta}$ is the probability of the refraction having a value
of $\theta$ arcsecs, and $\sigma$ is the rms refraction 'noise'.
\item the telescope beam can also be defined as a Gaussian:
  \begin{equation}
  S(\theta) = S_{o}\,exp\left[-4ln(2)\frac{\theta^2}{\theta^{2}_{fwhm}}\right]
  \label{eq:beam}
  \end{equation}
where S is the signal level which would be measured for a movement of the
telescope of $\theta$ arcsecs, and $\theta_{fwhm}$ is the FWHM size of the
beam in arcsecs.
\end{enumerate}

The expectation value of the
signal level, relative to the no refraction noise case, is given by
  \begin{equation}
  <S> = \frac{\int_{0}^{S_{o}}~S~P_{S}~dS}{\int_{0}^{S_{o}}~P_{S}~dS}
  \label{eq:expect}
  \end{equation}
where P$_{S}$ is the probability distribution for the signal. \\
Probabilities transform directly so that
  \begin{equation}
  P_{S}~dS = P_{\theta}(-d\theta)
  \label{eq:prob}
  \end{equation}
where use is made of the fact that the probability of S decreasing is a
function of the probability of $\theta$ increasing.\\
For simplicity, we rewrite (\ref{eq:beam}) as
  \begin{equation}
  S(\theta) = S_{o}exp\left[-\frac{\theta^2}{2\beta^{2}}\right]
  \label{eq:newbeam}
  \end{equation}
where $\beta^{2}$ = $\frac{\theta^{2}_{fwhm}}{8ln2}$~.
\\
Now, from equation (\ref{eq:newbeam}),
  \begin{equation}
  \frac{dS}{d\theta} = -\frac{\theta}{\beta^{2}}S
  \label{eq:6}
  \end{equation}
using this and equations (\ref{eq:rayleigh}),(\ref{eq:newbeam}),
  \begin{equation}
  P_{S} =
\frac{\beta^{2}}{\sigma^{2}}\frac{1}{S}\left[\frac{S}{S_{o}}\right]^\frac{\beta^{2}}{\sigma^{2}}
  \label{eq:7}
  \end{equation}
It should be noted that, if $\beta^{2}$ = $\sigma^{2}$, the probability is
uniform.
\\
If we now evaluate (\ref{eq:expect}) using (\ref{eq:7}) we find that
  \begin{equation}
  \frac{<S>}{S_{o}} = \frac{1}{1+\frac{\sigma^{2}}{\beta^{2}}} =
\frac{1}{1+\frac{8ln2\sigma^{2}}{\theta^{2}_{fwhm}}}
  \label{eq:8}
  \end{equation}
In the case of a perfect 15-metre telescope, one can write  for
$\theta_{fwhm}$ in arcsecs :
  \begin{equation}
  \theta_{fwhm} = [14.025 + 0.1856T_{E}]\lambda
  \label{eq:9}
  \end{equation}
  where $T_{E}$ is the illumination edge taper in decibels and $\lambda$ is
  the wavelength in mm \citep{gold87}. The edge taper determines the level of
  ground radiation accepted by the feed, and at JCMT $T_{E}\sim7.5$\,dB .  In
  the case of a telescope with surface imperfections one must use the measured
  value of the full width half-maximum beam size in (\ref{eq:8}).
  \\\\
  If we assume a 14.2$''$ beam at 850\,\micron{} and a 7$''$ beam at
  450\,\micron{}, then (\ref{eq:8}) becomes
  \begin{equation}
  \frac{1}{1~+~0.0275\sigma^{2}} ~ ~ {\rm for ~ the ~ 14.2" ~ beam}
  \label{eq:10}
  \end{equation}
and
  \begin{equation}
  \frac{1}{1~+~0.1132\sigma^{2}} ~ ~ {\rm for ~ the ~ 7" ~ beam}
  \label{eq:11}
  \end{equation}
where $\sigma$ is the rms refraction noise in arcseconds.

\label{lastpage} 

\begin{thebibliography}{28}
\expandafter\ifx\csname natexlab\endcsname\relax\def\natexlab#1{#1}\fi

\bibitem[{{Allan}(1966)}]{allan66}
{Allan} D., 1966, Proc. IEEE, 54, 221

\bibitem[{{Borys} {et~al.}(1999){Borys}, {Chapman}, \& {Scott}}]{borysnoise}
{Borys} C., {Chapman} S.~C., {Scott} D., 1999, \mnras, 308, 527

\bibitem[{{Church} \& {Hills}(1990)}]{churchhills90}
{Church} S., {Hills} R., 1990, in URSI/IAU Symposium on Radio Astronomical
  Seeing, pp. 75--80

\bibitem[{{Church} {et~al.}(1993){Church}, {Lasenby}, \& {Hills}}]{church93}
{Church} S.~E., {Lasenby} A.~N., {Hills} R.~E., 1993, \mnras, 261, 705

\bibitem[{{Coulson}(2001)}]{flydips}
{Coulson} I.~M., 2001. Tech. Rep. SCD/SN/005, JCMT,
  (http://www.jach.hawaii.edu/JACdocs/JCMT/SCD/SN/005)

\bibitem[{{Davis} {et~al.}(1997){Davis}, {Naylor}, {Griffin}, {Clark}, \&
  {Holland}}]{pwv}
{Davis} G.~R., {Naylor} D.~A., {Griffin} M.~J., {Clark} T.~A., {Holland} W.~S.,
  1997, ICARUS, 130, 387

\bibitem[{{Duncan}(1983)}]{duncan83}
{Duncan} W.~D., 1983, Infrared Physics, 23, 333

\bibitem[{{Duncan} {et~al.}(1995){Duncan}, {Robson}, {Ade}, \&
  {Church}}]{duncanchop}
{Duncan} W.~D., {Robson} I., {Ade} P.~A.~R., {Church} S.~E., 1995, in ASP Conf.
  Ser. 75: Multi-Feed Systems for Radio Telescopes, p. 295

\bibitem[{{Duncan} {et~al.}(1990){Duncan}, {Sandell}, {Robson}, {Ade}, \&
  {Griffin}}]{ukt14paper}
{Duncan} W.~D., {Sandell} G., {Robson} E.~I., {Ade} P.~A.~R., {Griffin} M.~J.,
  1990, \mnras, 243, 126

\bibitem[{{Goldsmith}(1987)}]{gold87}
{Goldsmith} P.~F., 1987, International Journal of Infrared and Millimetre
  Waves, 8, 771

\bibitem[{{Hazell}(1991)}]{hazellthesis}
{Hazell} A.~S., 1991, PhD thesis, Queen Mary and Westfield College

\bibitem[{{Holland} {et~al.}(2000){Holland}, {Duncan}, {Kelly}, {Peacocke},
  {Robson}, {Irwin}, {Hilton}, {Rinehart}, {Ade}, \& {Griffin}}]{scuba2holland}
{Holland} W.~S., {Duncan} W.~D., {Kelly} B.~D., {Peacocke} T., {Robson} E.~I.,
  {Irwin} K.~D., {Hilton} G., {Rinehart} S., {Ade} P.~A.~R., {Griffin} M.~J.,
  2000, in American Astronomical Society Meeting, Vol. 197, p. 5301

\bibitem[{{Holland} {et~al.}(1999){Holland}, {Robson}, {Gear}, {Cunningham},
  {Lightfoot}, {Jenness}, {Ivison}, {Stevens}, {Ade}, {Griffin}, {Duncan},
  {Murphy}, \& {Naylor}}]{scubapaper}
{Holland} W.~S., {Robson} E.~I., {Gear} W.~K., {Cunningham} C.~R., {Lightfoot}
  J.~F., {Jenness} T., {Ivison} R.~J., {Stevens} J.~A., {Ade} P. A.~R.,
  {Griffin} M.~J., {Duncan} W.~D., {Murphy} J.~A., {Naylor} D.~A., 1999,
  \mnras, 303, 659

\bibitem[{{Jenness} {et~al.}(1998){Jenness}, {Lightfoot}, \&
  {Holland}}]{remskypaper}
{Jenness} T., {Lightfoot} J.~F., {Holland} W.~S., 1998, \procspie, 3357, 548

\bibitem[{{Jenness} {et~al.}(2001){Jenness}, {Stevens}, {Archibald},
  {Economou}, {Jessop}, \& {Robson}}]{timfcfpaper}
{Jenness} T.~J., {Stevens} J.~A., {Archibald} E.~N., {Economou} F., {Jessop}
  N.~E., {Robson} E.~I., 2001, MNRAS, submitted

\bibitem[{{Le Poole} \& {van Someren Greve}(1998)}]{dreammode}
{Le Poole} R.~S., {van Someren Greve} H.~W., 1998, in Proc. SPIE Vol. 3357, p.
  638-643, Advanced Technology MMW, Radio, and Terahertz Telescopes, Thomas G.
  Phillips; Ed., Vol. 3357, p. 638

\bibitem[{{Masson}(1991)}]{masson91}
{Masson} C.~R., 1991, in ASP Conf. Ser. 19: IAU Colloq. 131: Radio
  Interferometry. Theory, Techniques, and Applications, p. 405

\bibitem[{{Naylor} {et~al.}(1994){Naylor}, {Clark}, \& {Davis}}]{naylorfts}
{Naylor} D.~A., {Clark} T.~A., {Davis} G.~R., 1994, \procspie, 2198, 703

\bibitem[{{Naylor} {et~al.}(2000){Naylor}, {Davis}, {Gom}, {Clark}, \&
  {Griffin}}]{naylorsubatmos}
{Naylor} D.~A., {Davis} G.~R., {Gom} B.~G., {Clark} T.~A., {Griffin} M.~J.,
  2000, \mnras, 315, 622

\bibitem[{{Olmi}(2001)}]{olmi2001}
{Olmi} L., 2001, A\&A, 374, 348

\bibitem[{{Omont} {et~al.}(1996){Omont}, {McMahon}, {Cox}, {Kreysa},
  {Bergeron}, {Pajot}, \& {Storrie-Lombardi}}]{omc96}
{Omont} A., {McMahon} R.~G., {Cox} P., {Kreysa} E., {Bergeron} J., {Pajot} F.,
  {Storrie-Lombardi} L.~J., 1996, \aap, 315, 1

\bibitem[{{Robson} {et~al.}(2001){Robson}, {Holland}, \&
  {Duncan}}]{scuba2robson}
{Robson} E.~I., {Holland} W.~S., {Duncan} W.~D., 2001, in UMASS/INAOE
  Conference `Millimetre Surveys for the Future', {Lowenthal} J., {Hughes}
  D.~H., eds., ASP Conf. Series, in press

\bibitem[{{Schieder} \& {Kramer}(2001)}]{sk01}
{Schieder} R., {Kramer} C., 2001, \aap, 373, 746

\bibitem[{{Smith} {et~al.}(2001){Smith}, {Naylor}, \& {Feldman}}]{irma}
{Smith} G., {Naylor} D., {Feldman} P., 2001, Int. J. of Inf. and Mm. Waves, in
  press

\bibitem[{{Stark}(2001)}]{southpole}
{Stark} A.~A., 2001, in Experimental Cosmology at Millimeter Wavelengths, {De
  Petris} M., {Gervasi} M., eds., astro-ph/0109229

\bibitem[{{Stevens} \& {Robson}(1994)}]{jasonsecant}
{Stevens} J.~A., {Robson} E.~I., 1994, MNRAS, 270, L75

\bibitem[{{Tilanus} {et~al.}(1997){Tilanus}, {Jenness}, {Economou}, \&
  {Cockayne}}]{remo97}
{Tilanus} R.~P.~J., {Jenness} T., {Economou} F., {Cockayne} S., 1997, in ASP
  Conf. Ser. 125: Astronomical Data Analysis Software and Systems VI, Vol.~6,
  p. 397

\bibitem[{{Wiedner} {et~al.}(2001){Wiedner}, {Hills}, {Carlstrom}, \&
  {Lay}}]{martina2001}
{Wiedner} M.~C., {Hills} R.~E., {Carlstrom} J.~E., {Lay} O.~P., 2001, \apj,
  553, 1036

\end{thebibliography}
\end{document}